\documentclass[sigconf,nonacm,review=false,screen,timestamp]{acmart}

\usepackage[english, ruled, linesnumbered]{algorithm2e}
\usepackage{multirow}
\usepackage{natbib}

\usepackage{listings}
\usepackage{tabularx}
\usepackage{graphicx,url}
\usepackage[utf8]{inputenc}

\def\tsc#1{\csdef{#1}{\textsc{\lowercase{#1}}\xspace}}
\tsc{WGM}
\tsc{QE}
\tsc{EP}
\tsc{PMS}
\tsc{BEC}
\tsc{DE}

\begin{document}

\title{SE$^3$M: A Model for Software Effort Estimation Using Pre-trained Embedding Models}                      
\author{Eliane M. De Bortoli F\'avero}
\orcid{0000-0002-8229-0296}
 \email{elianedb@utfpr.edu.br}
 \affiliation{%
   \institution{UTFPR - Federal University of Technology Paran\'a}
   \streetaddress{Via do Conhecimento, km 01}
   \city{Pato Branco}
   \state{PR}
   \country{Brazil}
   \postcode{85503-390}
 }
 
 \author{Dalcimar Casanova}
 \orcid{0000-0002-1905-4602}
 \email{dalcimar@utfpr.edu.br}
 \affiliation{%
   \institution{UTFPR - Federal University of Technology Paraná}
   \streetaddress{Via do Conhecimento, km 01}
   \city{Pato Branco}
   \state{PR}\
   \country{Brazil}
   \postcode{85503-390}
 }
 
  \author{Andrey Ricardo Pimentel}
 \orcid{}
 \email{andrey@inf.ufpr.br}
 \affiliation{%
   \institution{UFPR - Federal University of Paran\'a}
   \streetaddress{Rua Evaristo F. F. da Costa, 418}
   \city{Curitiba}
   \state{PR}\
   \country{Brazil}
   \postcode{85501-000}
 }

\begin{abstract}
Estimating effort based on requirement texts presents many challenges, especially in obtaining viable features to infer effort. Aiming to explore a more effective technique for representing textual requirements to infer effort estimates by analogy, this paper proposes to evaluate the effectiveness of pre-trained embeddings models. For this, two embeddings approach, context-less and contextualized models are used. Generic pre-trained models for both approaches went through a fine-tuning process. The generated models were used as input in the applied deep learning architecture, with linear output. The results were very promising, realizing that pre-trained incorporation models can be used to estimate software effort based only on requirements texts. We highlight the results obtained to apply the pre-trained BERT model with fine-tuning in a single project repository, whose value is the Mean Absolute Error (MAE) is 4.25 and the standard deviation of only 0.17, which represents a result very positive when compared to similar works. The main advantages of the proposed estimation method are reliability, the possibility of generalization, speed, and low computational cost provided by the fine-tuning process, and the possibility to infer new or existing requirements.

\end{abstract}

\keywords{
Software effort estimation, pre-trained model, context-less embedding, contextualized embedding, domain-specific model, BERT}

\maketitle

\section{Introduction}

Estimating software effort is a challenging and important activity in the software development process. This activity depends on the success of other crucial aspects of a project, mainly related to the achievement of time and budget constraints, directly impacting on the quality of the software product developed. The success of any particular software project depends heavily on how accurate its effort estimates are \cite{idri2016systematic}. An accurate estimate assists in contract negotiations, scheduling, and synchronization of project activities and efficient allocation of resources.

The importance of the accuracy of estimates has explored by studies in the field of software engineering (SE) published in recent years, such as:  \cite{kazemifard2011fuzzy},  \cite{azzeh2011adjusted}, \cite{bardsiri2013lmes}, \cite{hamdy2014improving}, \cite{amazal2014improving}, \cite{moharreri2016cost} and \cite{ionescu2017approach}. These studies continually seek to explore computational techniques individually or in combination, always seeking to achieve better levels of precision for effort estimation techniques.

Among the existing classifications for software effort estimation techniques, we highlight in this article non-algorithmic models - which do not use predefined metrics, such as points per function and points per use cases. These models make use of Machine Learning techniques (e.g. linear regression, neural networks) and are also called models by analogy \cite{shepperd1996effort}. According to the authors, in some ways, this method is a systematic way of judging specialists, who are already experts in seeking similar situations to inform their opinions. Manikavelan et al. \cite{manikavelan2014software} says that these models are built from historical data from projects to generate custom models.

According to Idri and Abran \cite{idri2016missing}, the accuracy of the estimate is improved when the analogy is combined with Artificial Intelligence (AI) techniques to generate estimates. In this way, fuzzy systems, decision trees, neural networks, and collaborative filtering are are some examples of techniques that improve Analogy-based Software Effort Estimation (ABSEE). Idri et al. \cite{idri2015analogy} reinforces that software effort estimation models by analogy are reproducible and closely resemble human reasoning because they based on experience gained from past projects.

ABSEE can fit either agile or traditional models as long as the estimation approach is based on data and previous team experiences to estimate software projects. One challenge that has been presented for using these techniques, especially in agile models, is the lack of project data and their requirements in the early stages of the development process. The basic specification of software requirements used in these models is the user story, which is user needs, usually written informally \cite{cohn2005agile}. 

Assigning effort estimates to software requirements, especially in the early stages of development, becomes quite critical as it depends on the empirical expertise of the experts involved (e.g. project managers and systems analysts), as there are not always complete records of historical data on projects and requirements. But this is still a limitation for most companies, as in most cases there are only textual requirements that briefly describe user needs. This fact makes this task very complex and sometimes even unfeasible. 

The difficulty with textual requirements (e.g. use cases, user stories) is related to intrinsic informality in many software development processes, which makes it difficult to use as a basis for predicting software costs. This limitation occurs because these texts include a diversity of domain-specific information such as natural language text containing source code, numeric data (e.g. IP addresses, artifact identification codes), among others. A very common aspect is the occurrence of different words, but they are used in the same context; in this case, they should be considered similars (polysemy) because their context is similar. Or, equal words (ambiguous), but applied in different contexts, therefore, should not be represented in the same way.

On the other hand, in the AI world, especially in the Natural Language Process (NLP) field, word embedding methods mainly aim to capture the semantics of a given word in a specific context. This method allows words to be represented densely and with low dimensionality, facilitating machine learning tasks that use textual characteristics. Breakthroughs in training word embedding models for a variety of purposes began in years recent with the emergence of Word2Vec \cite{mikolov2013efficient} and GloVe \cite{pennington2014glove}, enabling models to learn from a very large corpus. Thus, contextual representations through embeddings models have been very useful in identifying context-sensitive similarity \cite{huang2012improving}, disambiguation of the meaning of the word \cite{bordes2012joint}, \cite{chen2014unified}, the induction of the meaning of the word \cite{kaageback2015neural}, lexical substitution by the creation of a generic embeddings model \cite{melamud2016role}, sentence complementation \cite{liu2015learning}, among others. 

Some studies have been conducted specifically in the field of SE, such as recommending domain-specific topics from Stack Overflow question tags \cite{chen2016mining}, recommending similar bugs \cite{yang2016combining}, sentiment analysis in SE \cite{calefato2018journal}, embedding model using Word2Vec for the SE domain \cite{efstathiou2018word}, ambiguity detection in requirements engineering \cite{ferrari2019nlp}, among others.

More specifically applied in the generation of software effort estimates, the use of embeddings is highlighted in the studies by \cite{ionescu2017approach} and \cite{choetkiertikul2018deep}. In the first case, Ionescu (2017) explores the use of word embeddings generated by a context-less approach (Word2Vec), aggregated with design attributes and textual metrics (e.g. TF-IDF), from which positive results were obtained. Choetkiertikul et al. \cite{choetkiertikul2018deep} it seeks to infer estimates from the text of user stories, which were given as input to a deep learning architecture, with an embedding layer as input. However, these initiatives face two main limitations, which make it difficult to solve in the specific field of SE. Are they:

\begin{enumerate}
\item \textbf{Domain-specific terms present their meanings changed according to the context in which they are used:} in this article, domain-specific terms are a set of words common to a specific area (e.g. software engineering, medicine), among which there are strong semantic relationships. Some studies were carried out, seeking to develop resources to facilitate textual representation in software engineering \cite{tian2014sewordsim, efstathiou2018word}, but no complete solution for the presentation of contexts. Still, regarding the context representation, the textual requirements are usually short, bringing an inherent difficulty in identifying the context to which they refer. This reality makes it difficult to extract significant characteristics from the analyzed texts, making the inference process difficult.

\item \textbf{Lack of domain-specific SE data to train smart models:} this aspect makes it very difficult to train deep neural networks, which tend to overfiting themselves in these small training data sets, not reaching generalization. This reality is no different for textual software requirements and becomes more critical when we need these texts to be accompanied by their labels, which should represent the effort required to implement them.

\end{enumerate}

To solve both problems, this work explores the use of contextualized pre-training embedding models (e.g. BERT \cite{devlin2019bert}) to infer an estimate of software effort from textual requirements. Pre-trained models present the concept of transfer learning \cite{howard2018universal}, or allow to solve these problems, because we can train or model on a generic dataset (solve a lack of data problems) and adjust it (solver o problem of the specific meaning of words in different contexts).

Although there has been a lot of research on the application of word embedding in various areas, so far, only \cite{ionescu2017approach} and \cite{choetkiertikul2018deep} have sought to apply embeddings to estimate or run software. No research has explored the application of contextualized pre-trained embedding models following the task of ABSEE.

In this way, this article \textbf{aims to present a model for the inference of effort estimates by analogy, both for existing and new software requirements, using contextualized pre-trained embeddings models, having as input the exclusive use of textual requirements (e.g. user stories), generated in the initial stage of development}. This approach was named Software Effort Estimation Embedding Model (S$E^3$M). With this model, greater precision is sought for this activity, since it has currently been carried out based on human empiric experience. This characteristic makes these estimates quite subjective.

As noted by Howard and Ruder \cite{howard2018universal}, even though deep learning models have achieved state-of-the-art in many NLP tasks, these models are trained from scratch, requiring large datasets, and days to converge. Thus, pre-trained embeddings models make it easy to perform NLP tasks related to SE, without the need for training from scratch and with a low computational cost. According to the authors, this is possible through the fine-tuning technique, eliminating the need for a representative corpus.The fine-tuning approach consists of changing minimal task-specific parameters and is trained on the downstream tasks by simply fine-tuning all pre-trained parameters \cite{devlin2019bert}. Most language representation models (e.g. Word2Vec, ELMo, OpenAI GPT) are classified as context-less, i.e. each token considers only words on the left or right as part of its context \cite{vaswani2017attention}. This makes fine-tuning approaches difficult, where it is relevant to incorporate the context bidirectionally, the reason for using BERT models. BERT is a contextual representation model that solves the one-way constraint mentioned earlier, which will be further explained in section 2.3.

Therefore, the approach also aims to infer software effort estimation using pre-trained embeddings models, with and without fine-tuning on a SE specific corpus. Thus, with the results of this article, we intend to answer the following research questions (RQ):

\begin{itemize}
\item RQ1. Does a generically pre-trained word embedding model show similar results with a software engineering pre-trained model?

\item RQ2. Would embedding models generated by context-less methods (i.e. Word2Vec) be effective as models generated by contextualized methods (i.e. BERT)?

\item RQ3. Are pre-trained embeddings models useful to a text-based software effort estimation?

\item RQ4. Are pre-trained embeddings models useful to a text-based software effort estimation, both on new and existing projects?

\item RQ5. Are the results found generalizable, aiming to generate estimates of effort between projects or companies?
\end{itemize}

It is worth mentioning that, unlike the more similar approaches (\cite{ionescu2017approach}, \cite{choetkiertikul2018deep}), the proposed approach proposes to be generalizable, that is, it should allow the generation of estimates between projects and/or between companies, both for new requirements, as for existing ones (e.g. maintenance). In addition, similar approaches that perform the estimation process through text representation, apply textual representation methods by embeddings without context, which disregard the actual context of each word.

The structure of this article to organize as follows. Session 2 presents the background of software effort estimation and word embeddings (context-less and contextualized), in sequence to present the related works. Session 4 presents the theoretical aspects necessary for the proposed approach. Then, in session 5 the results of each step of the proposed method are presented, ending with session 6, where the initial research questions are to answer before concluding and presenting the future works.

\section{Background}

In this section, we first introduce aspects related to software effort estimation (Section 2.1). The following are the concepts of word embedding and the context-less and contextualized paradigms (Sections 2.2 and 2.3).

\subsection{Software Effort Estimation}

Various classifications for software effort estimation models have been applied in the last decades, with small differences, according to each author's point of view. According to Shivhare \cite{shivhare2014software}, software effort estimation models can subdivide into algorithmic/parametric and non-algorithmic. The first ones are those that use algorithmic models, applied to project attributes and/or requirements to calculate their estimate, presenting themselves as reproducible methods in substitution to non-algorithmic human expert methods \cite{kocaguneli2010use}. Examples of algorithmic models are COCOMO II \cite{boehm2000software} and Function Point Analysis \cite{choi2011arule}. Non-algorithmic ones are those based on Machine Learning techniques (e.g. linear regression, neural networks) and are also called models by analogy, which rely on historical data to generate custom models that can learn from this data \cite{manikavelan2014software}.

Chiu and Huang \cite{chiu2007adjusted} point out that the ABSEE estimate deals with the process of identifying one or more historical projects similar to the target project, and from them infer the estimate. In other words, but using the same line of reasoning, Shepperd's \cite{shepperd1996effort}, says that the basis for the ABSEE technique is to describe (in terms of several variables) the project that must be estimated, and then, to use this description to find other similar projects that have been completed. In this way, the known effort values for these completed projects can be used to create an estimate for the new project. 

Therefore, the ABSEE is classified as non-algorithmic. Idri and Abran \cite{idri2016systematic} also classify a technique by analogy as a machine learning technique. These authors further point out that machine learning models have also gained significant attention for effort estimation purposes, as they can model the complex relationship between effort and software attributes (cost factors), especially when this relationship is not linear, and it does not appear to have any predetermined form. Analog-based reasoning approaches have proven to be promising in the field of software effort estimation, and their use has increased among software researchers \cite{idri2016missing}.

Idri and Abran \cite{idri2016systematic} conducted a systematic review of the literature on ABSEE and found that these techniques outperform other prediction techniques. This conclusion to support by most of the works selected in their mapping. Among the main advantages of ABSEE is the similarity with human reasoning by analogy and, therefore, they are easier to understand.

Thus, the estimation of software effort by analogy is perceived as a very appropriate technique when the input resources are requirements specifications in the unstructured text format, as is the case of this research. Since it is possible to submit textual characteristics, drawn from these specifications, as input to machine learning models (e.g. neural networks).

\subsection{Word Embeddings}

Unlike lexical dictionaries (e.g. WordNet), which basically consist of a thesaurus, grouping words based on their meanings \cite{fellbaum2012wordnet} and which are usually built with human support, a word embedding model is made up of word representation vectors. Through these vectors, it is possible to identify a semantic relationship between words in a given domain, based on the properties observed in a training body and their automatically created generation \cite{mikolov2013efficient}.

Word embeddings are currently being a strong trend in NLP. Word embedding models use neural networks arquitectures to represent each word as a dense vector with low dimensionality and focusing on the relationship between words \cite{mikolov2013efficient}. Such vectors are used independently to calculate similarities between terms and as a basis of representation for NLP tasks (e.g. text classification, entity recognition, sentiment analysis). The word embedding models emerged to solve some limitations imposed by the bag-of-words (BOW) model, which usually present sparse and high dimensionality matrices.

Word2Vec, one of the most popular methods for generating word embeddings from a text corpus, is an unsupervised learning algorithm that automatically attempts to learn the relationship between words by grouping words that have similar meanings into similar clusters \cite{raschka2015python}. In the Word2Vec model \cite{mikolov2013efficient}, a neural network is trained to represent each word in the vocabulary as an n-dimensional vector. The general idea is that the distance between the vectors representing the word embedding is smaller if the corresponding words are semantically more similar (or related). Pennington et al. \cite{pennington2014glove} adds that for the generation of these vectors, the method captures the distributional semantics and co-occurrence statistics for each word in the presented training corpus.

The Word2Vec model internally implements two neural network-based approaches: Common Bag of words (CBOW) and skip-gram. Both are models for word embedding widely used in information retrieval tasks. The goal of the CBOW model is to predict the current word based on its context, while the skip-gram model is intended to predict the current surrounding words for the given word.
For both cases, the model consists only of a single weight matrix (in addition to the word analyzed), which results in training capable of capturing semantic information \cite{mikolov2013efficient}. After training, each word is to map to a low dimension vector. Results presented by their authors \cite{mikolov2013efficient} show that words with similar meanings are much closer to those with different meanings. Generally speaking, the key concept of Word2Vec is to find words that share common contexts in the training corpus, close to the vector space compared to others.

\begin{figure*}[ht]
  \centering
  \includegraphics[width= 1.0\textwidth]{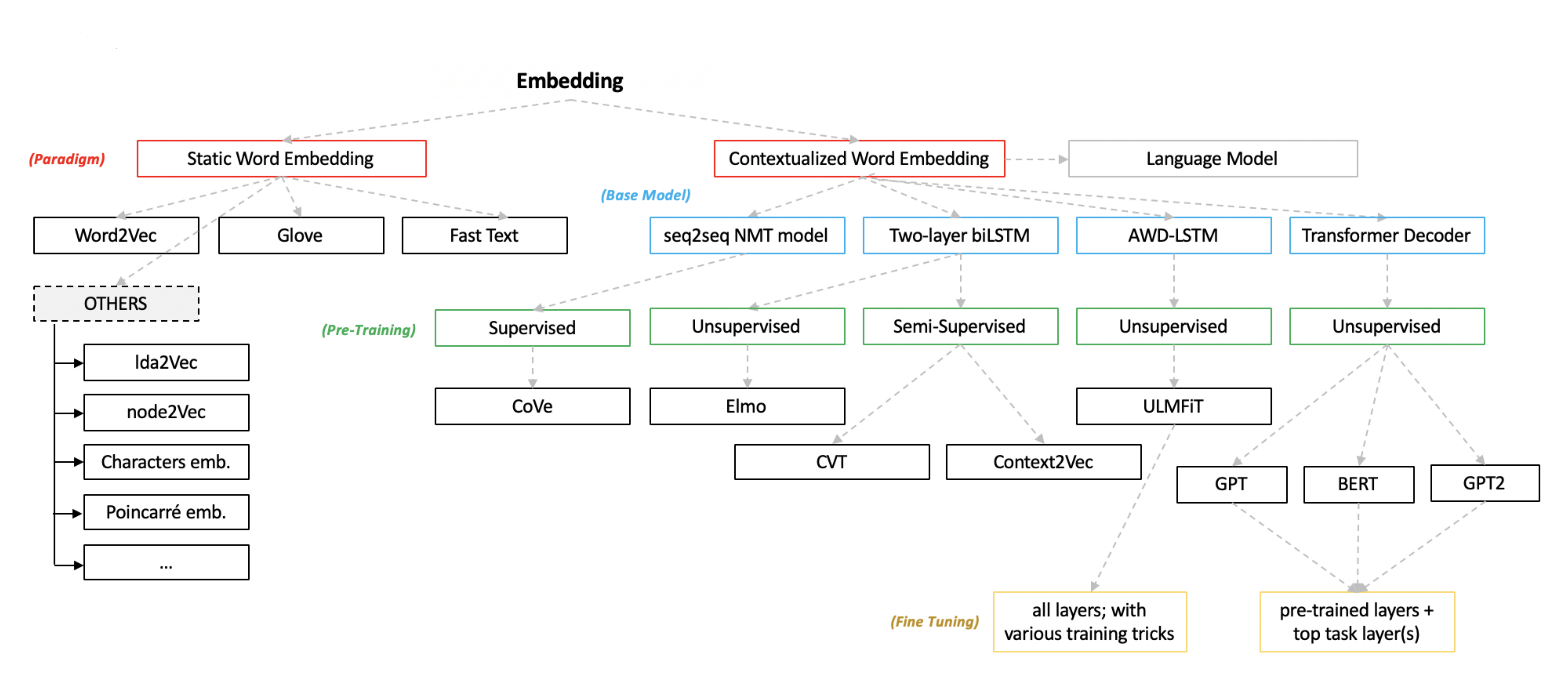}
  \caption{Differentiation between contextualized and context-less (static) embedding models (adapted of Haj-Yahia et al., 2019)} 
  \label{fig:figure0}
\end{figure*}

Word2Vec, like other models (e.g. Glove, FastText) is considered a context-less (static) method for generating pre-trained textual representations. This feature means that these models have constraints regarding the representation of the context of words in a text, making sentence-level or even fine-tuning token tasks difficult. Also, according to their authors \cite{mikolov2013efficient}, these models are to consider too shallow, as they represent each word by only one layer, and there is a limit to how much information they can capture. And finally, these models do not consider word polysemy, that is, the same word to use in different contexts can have different meanings, which is not dealt with by these models,

The Figure \ref{fig:figure0} presents a non-exhaustive differentiation between contextualized and context-less models. As we can see, Word2Vec is a form of static word embeddings such as Glove \cite{pennington2014glove}, Fast Text \cite{bojanowski2017enriching}, among others.

\subsection{BERT}

The BERT is an innovative method, considered the state of the art in pre-trained language representation \cite{devlin2019bert}. BERT models are considered contextualized or dynamic models, and have shown much-improved results in several NLP tasks \cite{dai2015semi}, \cite{peters2018deep}, \cite{radford2018improving}, \cite{howard2018universal} as sentiment classification, calculation of semantic tasks of textual similarity and recognition of tasks of textual linking.

This model originated from various ideas and initiatives aimed at textual representation that have emerged in the area of NLP in recent years, such as: coVe \cite{mccann2017learned}, ELMo \cite{peters2018deep}, ULMFiT \cite{howard2018universal}, CVT \cite{clark2018semi}, context2Vec \cite{melamud2016context2vec}, the OpenAI transformer (GPT and GPT-2) \cite{radford2018improving} and the Transformer \cite{vaswani2017attention}. 

BERT is characterized as a dynamic method, mainly because it has an attention mechanism, also called Transformer \cite{devlin2019bert}, which allows analyzing the context of each word in a text individually, including checking if each word has been previously used in a text with the same context. This allows the method to learn contextual relationships between words (or subwords) in a text.

BERT consists of several Transformer models \cite{vaswani2017attention} whose parameters are pre-trained on an unlabeled corpus like Wikipedia and BooksCorpus \cite{zhu2015aligning}. It can say that for a given input sentence, BERT “looks left and right several times” and outputs a dense vector representation for each word. For this reason, BERT is classified as a profoundly two-way model because it learns two representations of each word, one on the right and one on the left, and this learning to repeat n times. These representations are concatenated to obtain a final representation to use in future tasks.

The preprocessing model adopted by BERT accomplishes two main tasks: masked language modeling (MLM) and next sentence prediction (NSP). In the MLM task, the authors argue \cite{devlin2019bert} that it is possible to predict a particular masked word from the context. For example, let's say we have a phrase: "I love reading data science articles." We want to train a contextualized language model. In this case, you need to replace "data" with "[MASK]". It is a token to indicate that it is missing. We will then train the model so that it can predict "date" as the missing token: "I love reading articles from [MASK] science".

This technique aims to make the model learn the relationship between words, improving the level of learning, avoiding a possible “vicious cycle”, in which the prediction of a word to base on the word itself. Devlin et al. \cite{devlin2019bert} used 15-20\% of words as masked words.

The task of NSP is to learn the relationship between sentences. As with MLM, given two sentences (A and B), we want to know if B is the next sentence after A in the corpus or if it would be any sentence.

With this, BERT combines the pre-training tasks of both tasks (MLM and NSP), making it a task-independent model. For this, their authors provided pre-trained models in a generic corpus but allowing fine-tuning. It means that instead of taking days to pre-workout, it only takes a few hours. According to the authors of BERT \cite{devlin2019bert}, a new state of the art has been achieved in all NLP tasks they have attempted (e.g. Question Answering (QA) and Natural Language Inference (NLI)).

\section{Related Works}

Performing a systematic mapping focusing on ABSEE, we found few studies, as (\cite{abrahamsson2011predicting}, \cite{hussain2013approximation}, \cite{moharreri2016cost}, \cite{choetkiertikul2018deep}, \cite{zhang2016esse}, \cite{ionescu2017approach}, \cite{ochodek2016functional}, \cite{ayyildiz2016case}), that obtain the effort estimate from text using texts.

It was observed in most of the studies presented the bag of words approaches are applied, considering word-level features (e.g. tf, tf-idf, part-of-speech tag), which to treated individually, that, is based on quantitative and qualitative data, not employing specific knowledge about the text structure of the requirements, ignoring aspects of context. Only two studies (\cite{ionescu2017approach}, \cite{choetkiertikul2018deep}) differ from these attributes, as they apply word embedding models, but none of them use pre-trained embeddings models.

Ionescu \cite{ionescu2017approach} proposed a machine learning-based method for estimating effort for software development, using as input the text of project management requirements and metrics. The authors applied an original statistical preprocessing method to try out better results. First, a custom vocabulary was made. It is done using the standard deviation of the effort of those requirements where each word appears in the training set. For each requirement, a percentage of your words is maintained based on this statistic. The resulting requirements are concatenated with available project metrics. A modified TF-IDF calculation was also used, and numerical data were produced to form a bag-of-words, which is used as input to a linear regression algorithm.

Choetkiertikul et al. \cite{choetkiertikul2018deep} proposed the use of deep learning. Two neural network models were to combine into the proposed deep learning architecture: The Long Short Term Memory (LSTM), which are long term memories and the recurrent highway network. The model is trainable from start to finish with raw input data that has only gone through a preprocessing step. The model learns from the story point estimated by previous projects to predict the effort of new stories. This proposal \cite{choetkiertikul2018deep} uses context-less word embeddings as input to the LSTM layer. As input data, the title and description of the requirements report were combined into a single sentence, where the description follows the title. 

The embeddings vectors generated in this first layer serve as input to the LSTM layer, which then generates a representation vector for the full sentence. It should be to note that this process of training the embedding layer and then the LSTM layer, to then generate the embedding vectors for each sentence, becomes computationally expensive. For this reason, the authors pre-train these layers, and only then make these models available for use. This sentence vector is then to feed into the recurrent highway network, which transforms the document vector several times before producing a final vector that represents each sentence. Finally, the sentence vectors undergo simple linear regression, predicting their effort.

The possible bottleneck of this approach is the difficulty to feeding the model with new data. This feedback would fine-tune models, making them increasingly accurate. This difficulty occurs because with each new insertion into the dataset, the pre-training process, and consequently, its cost needs to be to repeat. Besides, Choetkirtkul's \cite{choetkiertikul2018deep} method realizes inter-project prediction, which is not repository independent.

Another aspect to point is that, because requirement texts do not usually have a structured form, more words may not represent more complexity \cite{zhang2016esse} and \cite{hussain2013approximation} and therefore greater effort. Actually, it is the context that influences the effort most.

As a differential, our visa method for making effort estimation in a generic way, that is, using a single requirements repository, independent of projects or repositories. A method of incorporating contextualized words (i.e. BERT) allows you to deeply learn the context of each selected word, solving problems of polysemy and ambiguity. When feeding the model with new training data, the adjustment can be made a few hours, that is, cheaper computationally than the generation of a pre-trained model from zero.

\section{Construction Approach}
\label{const_approach}

The main objective of this research is to evaluate the efficacy of pre-trained embedding models contextualized according to effort estimation and based on requirement texts. 

A requirement in this context can be a case of use, a user’s story, or any software requirement, provided that the data is in text format, and aligned with the target effort. For this paper, user’s stories serve as the input to the proposed model, and are composed of their description and their effort, provided in points per story. This data’s textual format requires some basic preprocessing (e.g. removal of special characters and stop words) before their use in the models.

It is important to note the nature of the software’s requirement texts, which are usually presented informally, that is, they do not have a standard format. In addition, these texts have a series of very specific elements (e.g. links, numeric addresses, method names). Considering these attributes, we propose that software automatically learn the characteristics of the original text.

\begin{figure}[!ht]
  \centering
  \includegraphics[width=.48\textwidth]{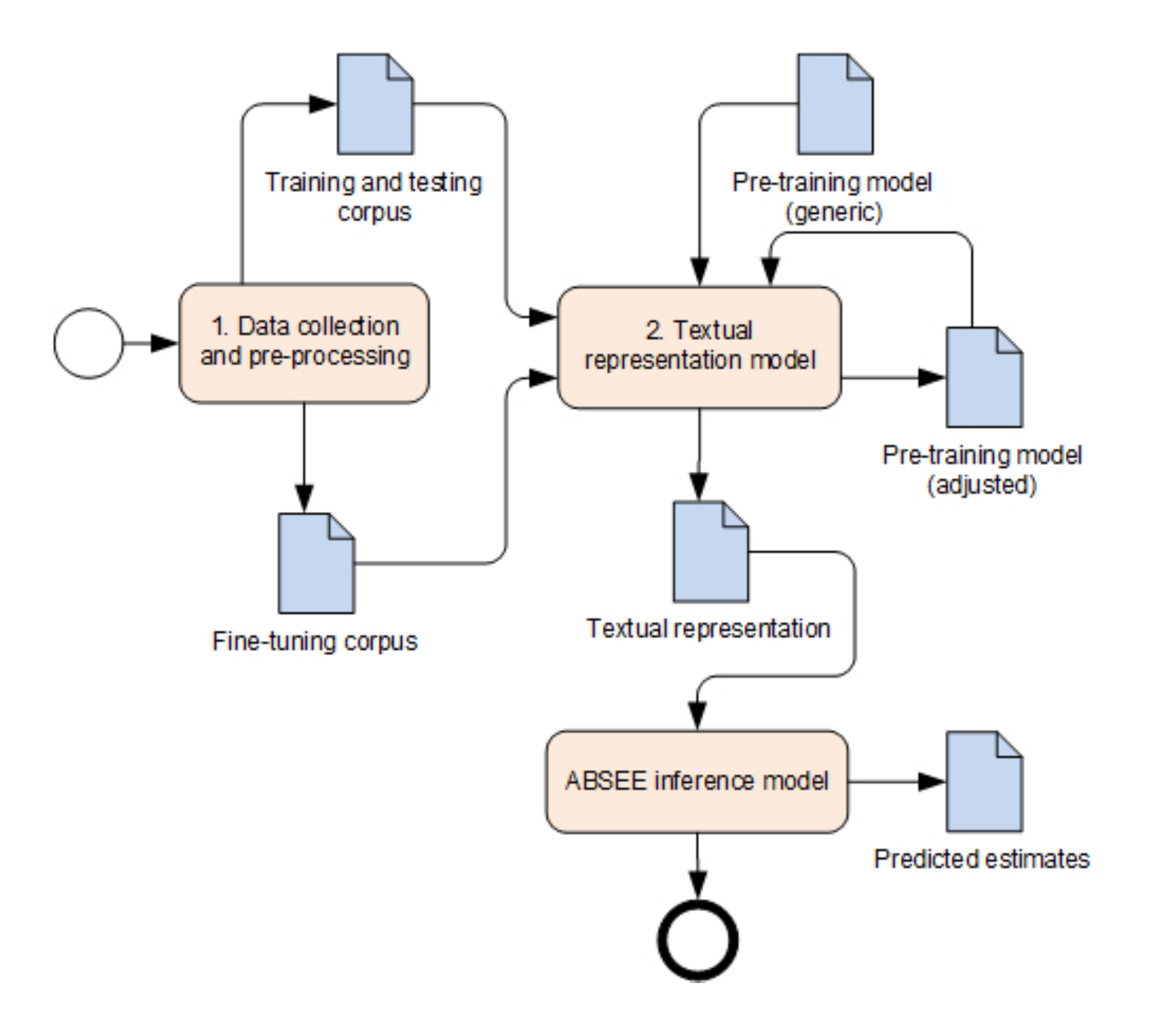}
  \caption{General architecture of the proposed model.} 
  \label{fig:figure1}
\end{figure}

For the purpose of a comparison, pre-trained models generated from two approaches for modeling language will be applied: context-less and contextualized models. Figure \ref{fig:figure1} presents the steps that make up the overall architecture of the proposed model, which will be described below.

\begin{enumerate}

\item \textbf{Data collection and pre-processing:} in this step, the data collection and preparation procedures are performed for later use during the \textit{feature learning} step, which will generate a context vector (i.e. numerical representation) for a given requirement text. For the proposed model, two corpus of texts containing software requirements will be required: \textit{corp\_SE} e \textit{corpPret\_SE} (as shown in Table \ref{tab:esp_corpus}). One of them will be the fine-tuning corpus, in which the texts are not labeled. The other corpus will be used during the training and testing stages of the inference model, in which each text will be labeled with their respective efforts. The texts for both corpus go through basic pre-processing procedures (e. g. removal of special characters and \textit{stopwords}).

\item \textbf{\textit{Textual representation model}:} this step consists in applying methods of deep learning to the textual characteristics \cite{bengio2012unsupervised}, aiming to generate the vector representation for each of the texts. Therefore, these methods do not use manual activity for the generation of characteristics. To achieve this goal, methods to generate context-less embeddings (e.g. Word2Vec) and contextualized embeddings (e.g. BERT) are used. Therefore, this step comprises the fine-tuning procedure of the generic pre-trained models for both approaches, which are given as input to the inference model. For fine-tuning, an unlabeled corpus (corpPret\_SE - according to Table \ref{tab:esp_corpus}) is applied together with generic pre-trained models for each incorporation approach (\textit{word2vec\_base} and \textit{BERT\_base} - according to Table \ref{tab:pret_model}). As an output, two pre-trained adjusted models are generated: \textit{word2vec\_SE} and \textit{BERT\_SE} (according to Table\ref{tab:model_ajuste}. Then, the pre-trained and adjusted models are used to extract the textual representations for each requirement that makes up the training and testing corpus, using appropriate pooling techniques (e.g. mean, sum) applied to embeddings for each word. This textual representation is given by a matrix containing the number of samples from the training and test corpuses in relation to the number of dimensions of the respective embeddings (context-less and contextualized). The vector representation of each requirement is given as an input to the ABSEE inference model, aiming to learn and infer new estimates.  

\item \textbf{Inference model for ABSEE}: the textual representation for the set of training and testing requirements is submitted as input to the inference model. This model is composed of a deep learning architecture, which is considered to be quite simplified, when compared to VGG-type models \cite{simonyan2014very}, Resnet \cite{he2016deep}, among others. It is important to note that the concept of deep learning is not related to the number of layers of neural networks that make up the architecture, but rather to the fact that this architecture executes a deep learning of the text’s characteristics, through the process of learning features, which begins by representing the texts using an embedding model. Therefore, the characteristics learned during this process are applied through the embedding layer of the deep learning architecture used. Since the network entry is a sequence of words (100 words are considered for each text), an LSTM layer has the function of processing that sequence, generating a representation for the sentence (that is, for each text). Subsequently, two dense layers with nonlinear activation functions are used (the dimensions of the hidden layers are 50 and 10 respectively), ending with a linear regression layer. This is an even smaller architecture, when compared to the one used in the work of \cite{choetkiertikul2018deep}, excluding the role of recurring networks \cite{graves2012supervised}, \cite{sak2014long} and applying a single sequence feedforward after the representation layer. This is possible due to the feature learning methods applied in the previous step. A simplified architecture also aims to not mask the results generated from the expected inputs to the network.

\end{enumerate}

After performing an inference of the estimates for each textual requirement for the training and testing sets, metrics for performance evaluation were applied, identified in the learning model used (according to the section \ref{sec:metric_eval}), in order to analyze the feasibility of the application of the model developed. This evaluation process was carried out using applicable statistical and graphic techniques, always related to the real development environment.

\subsection{Evaluation Metrics}
\label{sec:metric_eval}

The evaluation metrics used to evaluate model performance, which refers to the distance between the test set values and the predicted values. For this, some metrics were selected, which have been recommended for the evaluation of regression-based software effort prediction models \cite{sarro2016multi},\cite{choetkiertikul2018deep}, \cite{raschka2015python}, They are: Mean Absolute Error (MAE), Median Absolute Error (MdAE) and Mean Square Error (MSE).

\begin{equation}
    MAE = \frac{1}{N}\sum_{i=1}^{N} |actual\_eff - estimated\_eff\textsubscript{i}|	
\end{equation}

Where N is the number of textual requirements (e.g. user stories) that make up the test suite used to evaluate model performance, actual\_eff is the current effort measure, and estimated\_eff\textsubscript{i} is the estimated effort measure for a given textual requirement i.
We also used the Median Absolute Error (MdAE), suggested as a more robust metric for large outliers \cite{choetkiertikul2018deep}. MdAE is defined as:

\begin{equation}
    MdAE = median{|actual\_eff\textsubscript{i} - estimated\_eff\textsubscript{i}|}	
\end{equation}

The Mean Squared Error (MSE) metric, represented by Equation x, was also applied:

\begin{equation}
    MSE = \sqrt{\frac{1}{N}\sum_{i=1}^{N} (actual\_eff\textsubscript{i} - estimated\_eff\textsubscript{i})^2}	
\end{equation}

\subsection{Data Collection and Pre-Processing}
\label{section4}

In order to obtain and prepare the data that makes up the training and testing corpus, and the fine-tuning corpus, the data collection and pre-processing step was performed using API's for the NLP, based on models previously established by the literature. These steps precede the process of representing textual characteristics, that is, the generation of the context vector for each requirement.

Thus, in order to create the training and testing data set, a corpus specific to the software engineering context (\textit{corp\_SE}) was used, composed of textual requirements, more specifically user’s stories \cite{choetkiertikul2018deep}, labeled according to their respective development efforts. It is important to highlight that, despite being referred to as user’s stories, the text with requirements does not have a standard structure. Regarding the effort attributed to each requirement, it is worth mentioning that no single measurement scale was adopted (ex. \textit{Fibonacci}). The \textit{corp\_SE} (Table \ref{tab:esp_corpus}) is considered by the authors to be the first data set where the focus is on the level of requirements (e.g. user’s stories) and not just on the project level, as in most data sets available for SE research.

The requirements texts, as well as the effort given to each of them, were obtained from large open sources from project management systems (e.g. Jira), totaling 23.313 requirements (Table \ref{tab:numreq_Project}), which were initially made available by \cite{porru2016estimating}. Subsequently, \cite{choetkiertikul2018deep} used the same database to carry out his research, aiming to estimate software effort by analogy. Despite the difficulty in obtaining the real effort to implement a software requirement, the authors claim to have been able to obtain the implementation time based on the situation (\textit{status}) of the requirement. Thus, the effort was obtained beginning from the moment when the situation was defined as "in progress" until the moment when it was changed to "resolved". Thus, \cite{choetkiertikul2018deep} applied two statistical tests (\textit{Spearman's} and \textit{Pearson} correlation) \cite{zwillinger1999crc}, which suggested a correlation between the points throughout the history and their real effort. Therefore, this same database was applied to the research proposed for this paper. It is known that these story points were estimated by human teams and, therefore, may contain biases and, in some cases, may not be accurate, which may cause some level of inaccuracy in the models.

\begin{table}[!ht] 
    \footnotesize
    \begin{tabularx} {0.48\textwidth}{l|X|c}
      \hline
      \textbf{Projct ID} & \textbf{Description} & \textbf{Requirements/project} \\
      \hline
      {0} & {Mesos} & {1680}\\
      {1} & {Usergrid} & {482}\\
      {2} & {Appcelerator Studio} & {2919}\\
      {3} & {Aptana Studio} & {829}\\
      {4} & {Titanium SDK/CLI} & {2251}\\
      {5} & {DuraCloud} & {666}\\  
      {6} & {Bamboo} & {521}\\
      {7} & {Clover} & {384}\\
      {8} & {JIRA Software} & {352}\\
      {9} & {Moodle} & {1166}\\
      {10} & {Data Management} & {4667}\\
      {11} & {Mule} & {889}\\
      {12} & {Mule Studio} & {732}\\
      {13} & {Spring XD} & {3526}\\
      {14} & {Talend Data Quality} & {1381}\\
      {15} & {Talend ESB} & {868}\\
      \hline
      \textbf{Total} & {} & \textbf{23.313}\\
      \hline
    \end{tabularx}
    
    \caption{Number of textual requirements (user stories) and description of each of the 16 projects used in the experiments \cite{choetkiertikul2018deep}.}
    \label{tab:numreq_Project}
\end{table}

Typically, user’s stories are measured on a scale based on a series of \textit{Fibonacci} \cite{scott2014origin}, called \textit{Planning Poker} (e.g. 1, 2, 3, 5, 8, 13, 21, 40, 100) \cite{cohn2005agile}. As there is no standardized use of this scale among the projects used to create \textit{corp\_SE}, there was no way of approximating the points by the history provided. Therefore, 100 possible predictions were considered, distributed over the \textit{corp\_SE}, of which some are nonexistent, as can be seen in the histogram of Figure \ref{figure3}. In this way, the effort estimate was treated as a regression problem.

\begin{table}[!ht] 
    \footnotesize
    \begin{tabularx}{0.48\textwidth}{l|X|c|c}
      \hline
      \textbf{Corpus} & \textbf{Specification} & \textbf{Aplication} & \textbf{Labeled}\\
      \hline
      \textit{corp\_SE} & Contains 23.313 user stories and consists of 16 large open source projects in 9 repositories (Apache, Appcelerator, DuraSpace, Atlassian, Moodle, Lsstcorp, Mulesoft, Spring e Talendforge). & Train and test & YES \\
      \hline
      \textit{corpPret\_SE} & It consists of more than 290 thousand texts of software requirements of different projects (ex. \textit{Apache, Moodle, Mesos}). & Fine-tuning & NO\\
      \hline
    \end{tabularx}
    
    \caption{Corpus used in the experiments carried out.}
    \label{tab:esp_corpus}
\end{table}

For the fine-tuning process, which makes up the proposed model, a corpus of specific texts from software engineering, the \textit{corpPret\_SE} (shown in the Table (\ref{tab:esp_corpus}), is used. It is not labeled, therefore the training carried out is unsupervised, and is composed of texts with specifications of requirements, obtained from open source repositories, according to the procedure described by \cite{choetkiertikul2018deep}.

\begin{figure}[!ht]
  \centering
  \includegraphics[width=.48\textwidth]{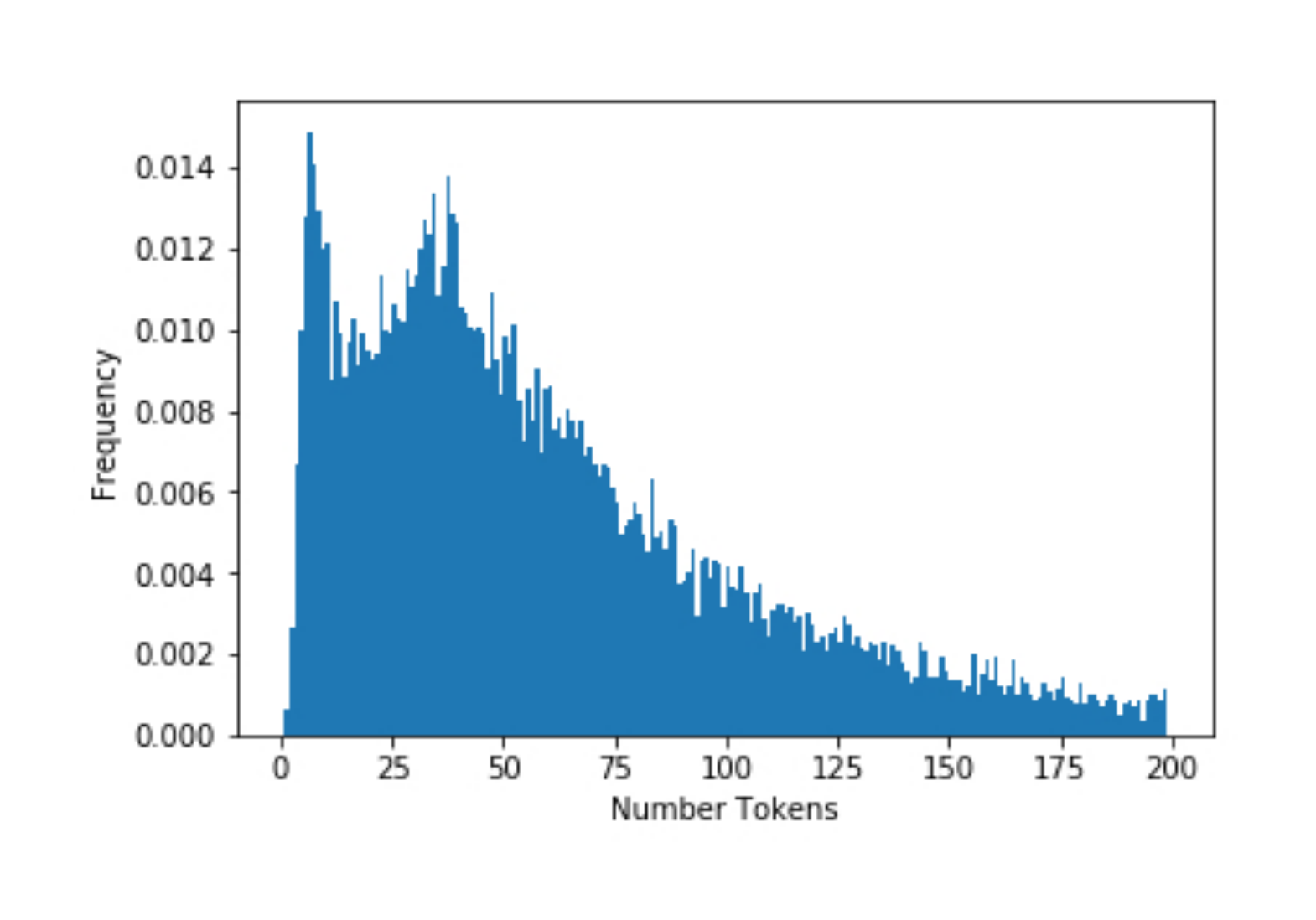}
  \caption{Histogram representing the number of words in each sentence of the dataset of training and testing.} 
  \label{figure2}
\end{figure}

\begin{figure}[!ht]
  \centering
  \includegraphics[width=.48\textwidth]{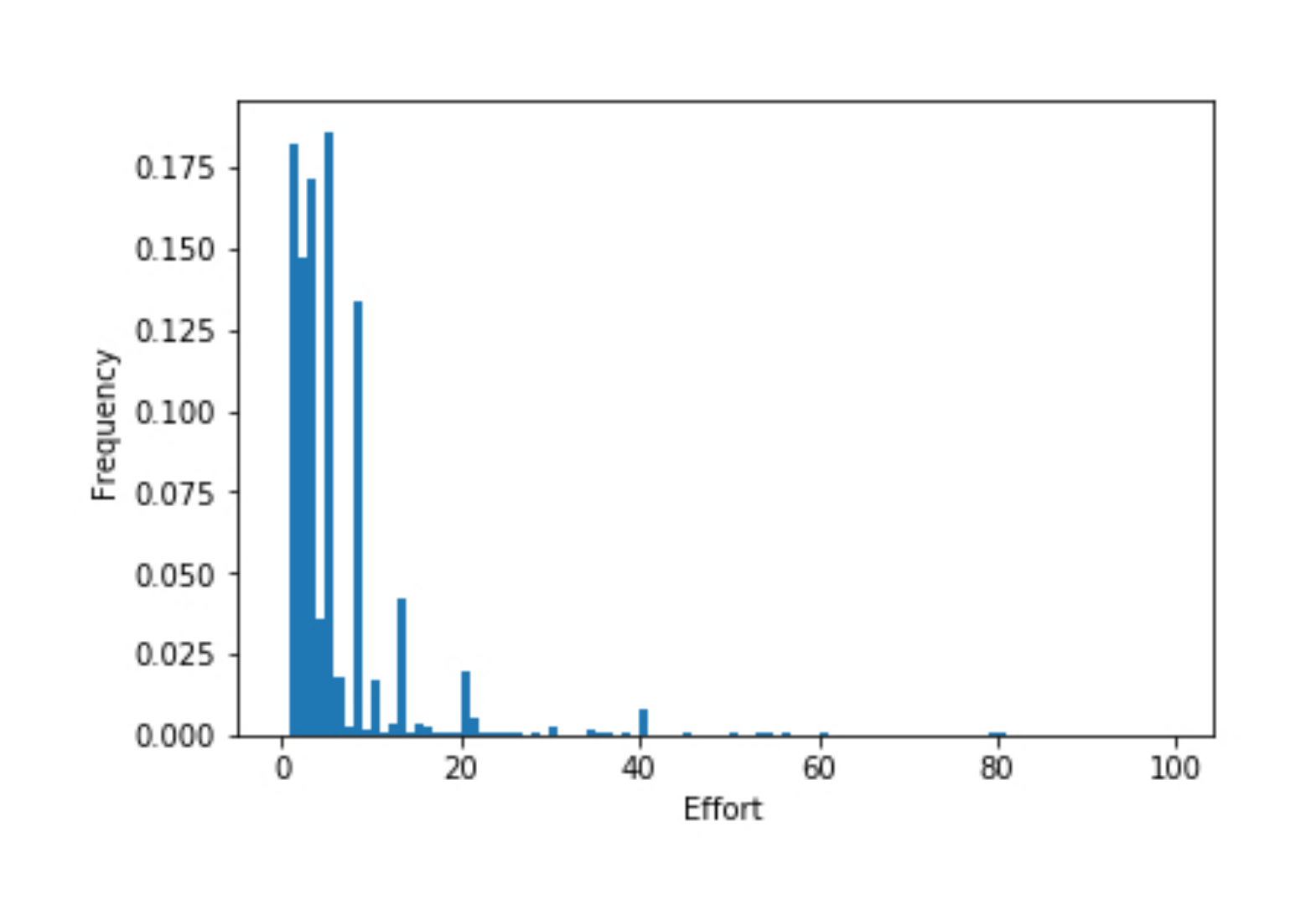}
  \caption{Histogram representing the size of the effort in relation to its frequency in the corp\_SE.} 
  \label{figure3}
\end{figure}

While exploring the data available in \textit{corp\_SE}, some relevant aspects that interfere with the inference model’s settings were observed. The histogram of Figure \ref{figure2} allows for one to evaluate the maximum number of words to be considered per text. The average number of words per text, accompanied by its standard deviation, is $53\pm108.6$.

Figure \ref{figure3} shows the frequency of distribution of the effort size in the training and testing database (\textit{corp\_SE}). It can be seen that most of the samples have smaller efforts (between 1 and 8 points per story). 

The cross validation \textit{k-fold} method was applied in order to partition the data set used to carry out the experiments (\textit {corp\_SE}). Thus, a number of equal subsets (\textit{nsplit}) was defined, represented by \textit{k} with a value of 10. Thus, the data from the \textit{corp\_SE} were divided into a set of training and validation (90\% of texts), and a set of tests (10\% of texts). For each subset of the data, the mean and standard deviation for the metrics applied in the performance evaluation were obtained, as described in the section \ref{sec:metric_eval}.

\subsection{Textual Representation Model}

The purpose of the procedures described in this section is to generate models to represent the texts that make up the training and test corpus. These representation models will be obtained from the generic and adjusted pre-trained \textit{embeddings} models, that must consider the diversity of existing contexts. Thus, the representation models (according to the Figure \ref{fig:figure1}) serve as input to the proposed sequential architecture.

To perform the experiments, two pre-trained generic \textit{word embeddings} models were applied, one using the context-less approach (\textit{Word2Vec}) and the other the contextualized approach (\textit{BERT}). 

Thus, for the context-less approach, a pre-trained model called \textit{word2vec\_base} was used, the specifications of which are shown in Table \ref{tab:pret_model}. As a contextualized model, a generic BERT model (\textit{BERT\_base uncased}) had previously been pre-trained and made available by its authors \cite{devlin2019bert} for free use in PLN tasks, as specified in Table\ref{tab:pret_model}.

\begin{table}[htp] 
    
    \footnotesize

    \begin{tabularx}{0.48\textwidth}{X|X}
      \hline
      \textbf{Pre-trained model} & \textbf{Specification}  \\
      \hline
      \textit{word2vec\_base} & trained on a \textit{corpus} from \textit{Wikipedia} using the \textit{Word2Vec} \cite{mikolov2013efficient} algorithm. For this, the following hyperparameters were used: number of dimensions of the hidden layer = 100; method applied to the learning task = \emph{cbow}. \\
      \hline
      \textit{BERT\_base} &  \textit{Bert\_base uncased}: 12 layers for each \textit{token}, 768 hidden layers, 12 heads of attention, 110 million parameters.
       
      The \textit{uncased} specification means that the text was converted to lower case before \textit{tokenization} based on \textit{WordPiece}, in addition, removes any accent marks. This model was trained with english texts (Wikipedia) with lowercase letters. \\
      \hline
    \end{tabularx}
    
    \caption{Pre-trained models used in the performed experiments.}
    \label{tab:pret_model}
\end{table}

The \textit{BERT\_base} model, as well as the other pre-trained BERT models, offers 3 components \cite{devlin2019bert}:

\begin{itemize}
\item A \textit{TensorFlow checkpoint (bert\_model.ckpt)} that contains pre-trained weights (consisting of 3 files).

\item A vocabulary file (vocab.txt) for mapping \textit{WordPiece} \cite{zhang2016google} for word identification.

\item A configuration file (\textit {bert\_config.json}) that specifies the model's hyperparameters.

\end{itemize}

Then, these two generic models go through a fine-tuning process, as presented in the section \ref{sec:fine-tuning}.

\subsection{Fine-tuning}
\label{sec:fine-tuning}

It is worth noting that the fine-tuning process consists of the use of a pre-trained embedding model (trained on a generic dataset) in an unsupervised way, which is adjusted, that is, retrained on a known data set that is specific to the area of interest. In this case, the fine-tuning was performed on the generic models \textit{word2vec\_base} and \textit{BERT\_base}, using corpus \textit{corpPret\_SE} (shown in Table \ref{tab:esp_corpus}). 

\begin{figure}[!htb]
\centering
\includegraphics[width=.48\textwidth]{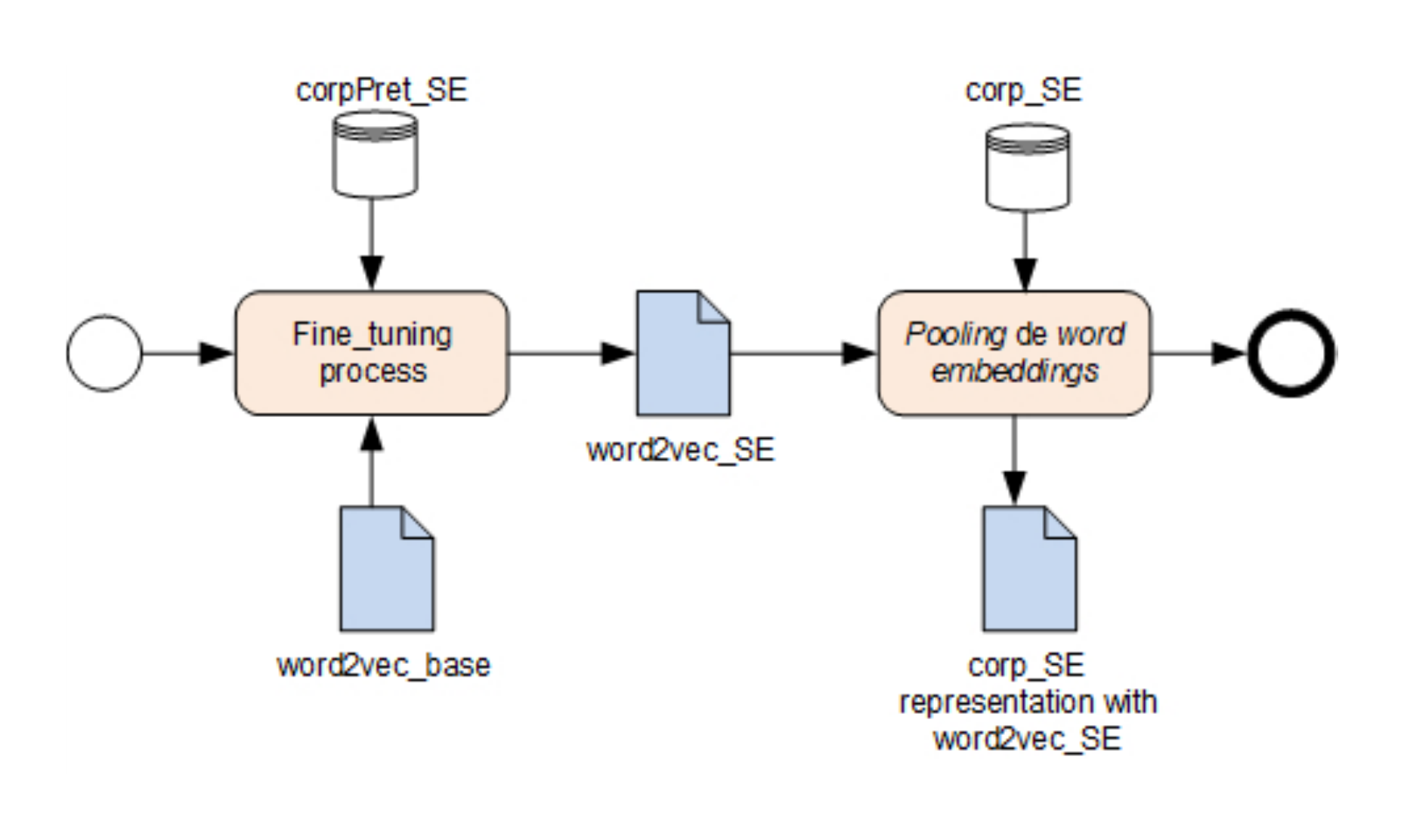}
\caption{Pipeline of the \textit{word2vec\_base} fine-tuning process and generation of textual representation for the corp\_SE.}
\label{fig:fig_preWord2vec}
\end{figure}

\begin{figure}[!htb]
\centering
\includegraphics[width=.48\textwidth]{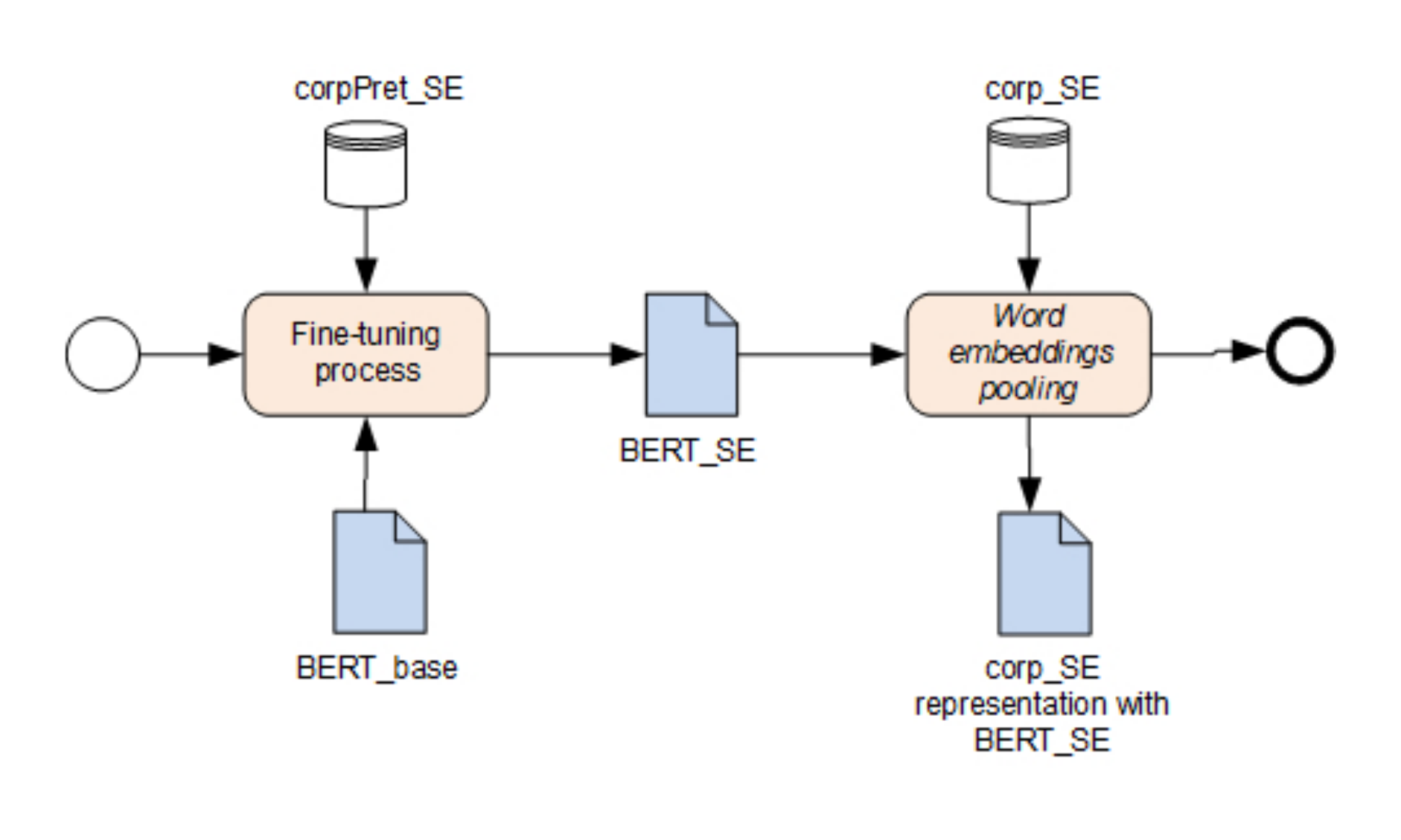}
\caption{Pipeline of the fine-tuning process of the \textit{BERT\_base} model and generation of the textual representation for the corp\_SE.}
\label{fig:fig_ajusteBert}
\end{figure}

The following are the pipelines (Figures \ref{fig:fig_preWord2vec} and \ref{fig:fig_ajusteBert}) used to adjust and generate the representation of the texts \textit {corp\_SE}. This representation was used as input to the proposed sequential architecture.

A fine-tuning of the generic model \textit{word2vec\_base} (Figure \ref{fig:fig_preWord2vec}) was performed using specific methods for this purpose provided by the \textit{Gensim} library in the \textit{Python} language. This process generated the \textit{word2vec\_SE} model. This adjusted model was used to generate the average representation (see section \ref{obtCaract}) for each requirement text of the training and testing corpus (corp\_SE).

The fine-tuning process of the pre-trained BERT model consists of two main steps \cite{devlin2019bert}:

\begin{enumerate}
    
\item \textbf{Preparation of data for pre-training:} initially the input data is generated for pre-training. This is done by converting the input sentences into the format expected by the BERT model (using \textit{create\_pretraining\_data} algorithm). As BERT can receive one or two sentences as input, the model expects an input format in which special tokens mark the beginning and end of each sentence, as shown in Table \ref{tab:exe_entradaBERT}. In addition, the \textit{tokenization} process needs to be performed. BERT provides its own \textit{tokenizer}, which generates output as shown in Table \ref{tab:exe_tokenizerBERT}.

\begin{table}[ht]
    
    \footnotesize
    \begin{tabularx}{0.48\textwidth}{X|X}
      \hline
      \textbf{Entry of two sentences} & \textbf{Entry of a sentence}  \\
      \hline
      [CLS] The man went to the store. [SEP] He bought a gallon of milk. [SEP] & [CLS] The man went to the store. [SEP] \\
      \hline
    \end{tabularx}
    
    \caption{Example of formatting input texts for pre-training with BERT.}
    \label{tab:exe_entradaBERT}
\end{table}

\begin{table}[ht]
    \footnotesize
    \begin{tabularx}{0.48\textwidth}{l|X}
      \hline
      \textbf{Input sentence} & "Here is the sentence I want embeddings for."  \\
      \hline
      \textbf{Text after \textit{tokenizer}} & ['[CLS]', 'here', 'is', 'the', 'sentence', 'i', 'want', 'em', '\#\#bed', '\#\#ding', '\#\#s', 'for', '.', '[SEP]'] \\
      \hline
    \end{tabularx}
    
    \caption{Example application of \textit{tokenizer} provided by BERT.} 
    \label{tab:exe_tokenizerBERT}
\end{table}

\item \textbf{Application of the pre-training method:} the method used for pre-training by BERT (\textit{run\_pretraining}) was made available by its authors. The necessary hyperparameters were informed, the most important being:
\begin{itemize}
\item \textit{input\_file}: directory containing pre-formatted pre-training data (as per step 1).

\item \textit{output\_dir}: output file directory.

\item \textit{max\_seq\_length}: defining the maximum size of the input texts (set at 100).

\item \textit{batch-size}: maximum lot size (set at 32, per use guidance of the pre-trained model \textit{BERT\_base}.

\item \textit{bert\_config\_file}: BERT model configuration file, supplied with the pre-trained model (\textit{bert\_config.json}).

\item \textit{init\_checkpoint}: files of the pre-trained model used containing the weights (\textit{bert\_model.ckpt}).

\end{itemize}

\end{enumerate}

For the generic BERT model (Table \ref{tab:pret_model}, we opted for its version \textit{Uncased\_L-12\_base}, here called \textit {BERT\_base} (Table \ref{tab:pret_model}). The fine-tuning process for the BERT model also used the \textit{corpPret\_SE} (Figure \ref{fig:fig_ajusteBert}) and was performed as shown above.

The entire process, from data preparation to fine-tuning the BERT model, used the algorithms produced in the repository \cite{BERTGit}, in which \cite{devlin2019bert} provides the full \textit {framework} developed in the \textit {Python} language.

After performing the fine-tuning, two new pre-trained models are available, as shown in Table \ref{tab:model_ajuste}, which will also compose the experiments.

It is noteworthy that the proposed model requires pre-training only for the embedding layer. This allows, for example, for this pre-trained model to be made available for other software engineering tasks, or even for different effort estimation tasks. Thus, this pre-trained model may undergo successive adjustments, according to the need of the task to which it will be applied.

\begin{table}[ht]
    \footnotesize
    \begin{tabularx}{0.48\textwidth}{l|X}
      \hline
      \textbf{Pre-trained models} & \textbf{Specification}  \\
      \hline
      \textit{word2vec\_SE} & consists of the \textit{word2vec\_base} model after fine-tuning with the corpus \textit{corpPret\_SE}. \\
      \hline
      \textit{BERT\_SE} &  consists of the \textit{BERT\_base} model after fine-tuning with the corpus \textit{corpPret\_SE}. \\
      \hline
    \end{tabularx}
    \caption{Adjusted pre-trained models applied to the performed experiments.}
    \label{tab:model_ajuste}
\end{table}

\subsection{Obtaining Characteristics} 
\label{obtCaract}

After the fine-tuning was completed, processing was performed to obtain textual representations from the four models of embeddings (\textit{word2vec\_base, BERT\_base, word2vec\_SE and BERT\_SE)}. For the context-less embeddings model, represented by the \textit{word2vec\_base} and \textit{word2vec\_SE} models (as shown in Figure \ref{fig:fig_preWord2vec}), the embeddings vectors of the words contained in each text were averaged \cite{wieting2017revisiting, palangi2016deep}.

As for the contextualized embeddings model, the textual representations generated by the BERT model (according to Figure \ref{fig:fig_ajusteBert}) present a different structure from the context-less embeddings models (e.g. Word2Vec). This is primarily due to the fact that the number of dimensions of the embeddings vectors is not defined by the user, but by the model itself. Therefore, the number of dimensions of word embeddings for the model  \textit{BERT\_base} is defined in 768. In addition, each word in this model is represented by 12 layers (standard for \textit{BERT\_base}), with the need to pool \cite{lev2015defense} the embeddings of some of the layers for each word. In order to define the pooling strategies to be applied, the article by \cite{devlin2019bert} was taken as a basis.

Thus, one of the proposed strategies was used, considering that there were no significant differences between the results obtained with the other tested strategies. Thus, for the models \textit{BERT\_base} and \textit {BERT\_SE}, the strategy chosen was to use the penultimate layer of each word to generate a vector of average embeddings for each sentence. More details on pooling strategies can be found in \cite{lev2015defense, alammar2018illustrated}.


\subsection{Exploratory data analysis}

Before presenting the results of the effort estimation, it is first important to highlight some observable aspects regarding the vector of textual representation obtained after the embeddings layer. These sentence embeddings were generated from the models specified in Tables \ref{tab:pret_model} and \ref{tab:model_ajuste}, that is, generic and fine-tuned models for both approaches (without context and contextualized).

In order to show the characteristic of the generated embeddings, the \textit{t-Distributed Stochastic Neighbor Embedding} (t-SNE) algorithm was applied. This algorithm  has been used to represent complex data graphically and in smaller dimensions, while preserving the relationships between neighboring words \cite{maaten2008visualizing}, which greatly facilitates their understanding.

\begin{figure}[!htb]
\centering
\includegraphics[width=.48\textwidth]{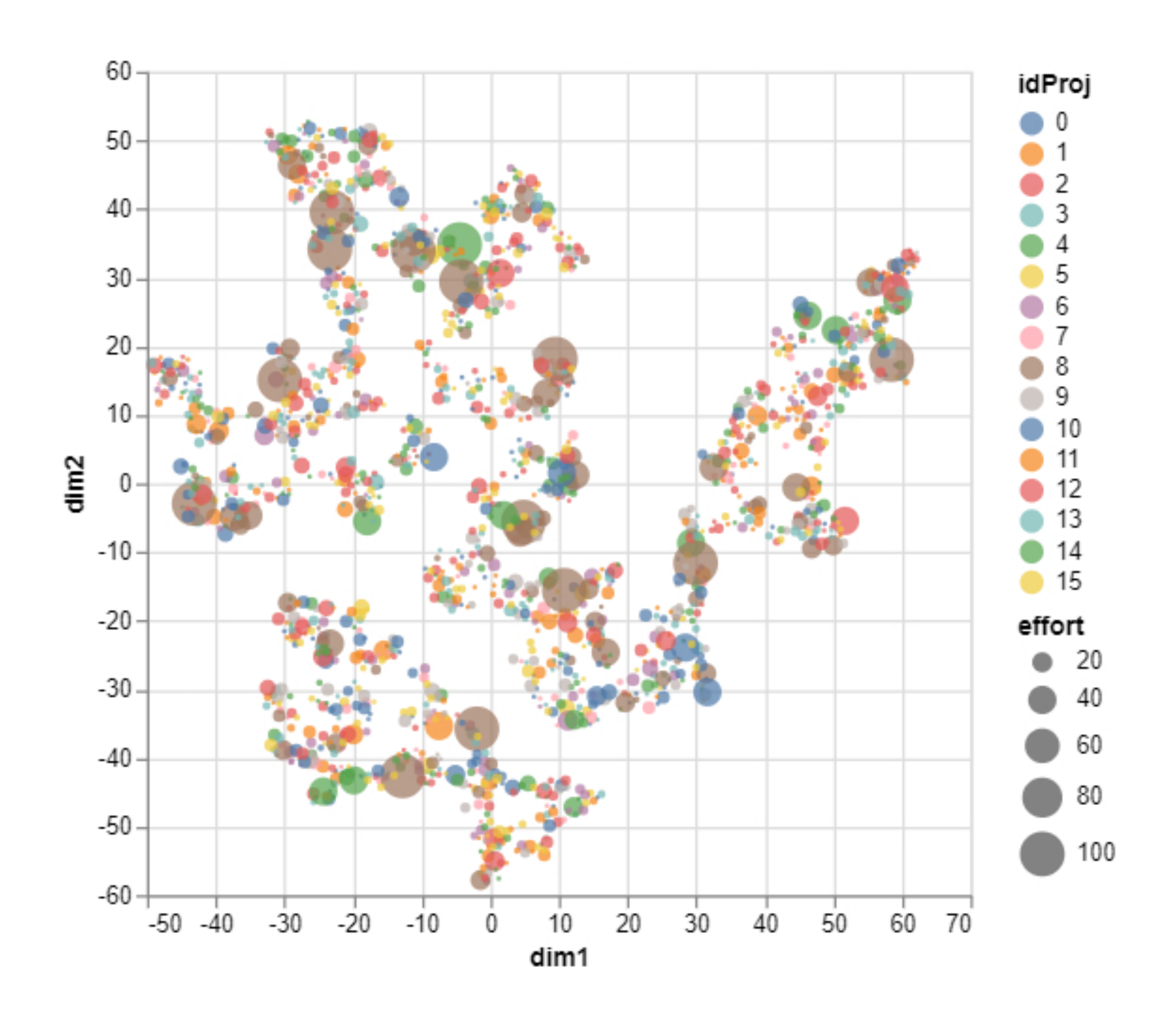}
\caption{Embeddings generated by \textit{BERT\_SE}. The points represent the effort for each requirement, according to its size. The larger the point size, the greater the effort.}
\label{fig:projIdEffort}
\end{figure}

Thus, Figure \ref{fig:projIdEffort} shows sentence embeddings generated from the \textit{BERT\_SE} for each textual requirement. Only 100 instances of requirements were included for each of the 16 projects analyzed (Table \ref{tab:numreq_Project}) (represented by “idProj”). The \textit{t\_SNE} algorithm reduced the 768-dimension model (BERT standard) to just two dimensions, which allowed for a visual analysis of the representations obtained and, based on these representations, some conclusions were reached. 

Table \ref{exe2_simTextos} present examples of requirement texts extracted from 2 different groups according to Figure \ref{fig:projIdEffort}. The texts, in each of the tables, have minimum distances between them.

\begin{table*}[ht]
    \footnotesize
    \begin{tabularx}{0.98\textwidth}{l|X}
      \hline
      \textbf{ID} & \textbf{Text}  \\
      \hline
      \textit{Texto 18994} & for the \textbf{sqlserver and postgresql connection} can not show the structure correctly.1. \textbf{create a\textbf{postgresql sqlserver connection} set} or not set the catalog parameter(doesn't set the textbf{schema}). 2. check the structure, only one \textbf{schema} show under each catalog. in fact i have several. please check it same issue as \textbf{informix db}. \\
      \hline
      \textit{Texto 17552} & sqoop - unable \textbf{to create job using merge command} as a user, i need to use xd \textbf{sqoop} module to support the merge command.  currently, the sqoop runner createfinalarguments method forces the requirement for \textbf{connect}, \textbf{username and password} options which are not valid for the merge option. a check of the module type to not force these options being assigned to \textbf{sqoop arg list} would be preferred. \\
      \hline
      \textit{Texto 15490} & need \textbf{to create a persistent-job-registry} in order to hook up the \textbf{to get access} to all the jobs available the job registry has to be shared.  currently the only \textbf{implementation} is the mapjobregistry. testability. the admin will need to be see all \textbf{jobs created} by its containers. \\
      \hline
    \end{tabularx}
    \caption{Contextual similarity between grouped texts.}
    \label{exe2_simTextos}
\end{table*}

In this example (Table \ref{exe2_simTextos}), the context of the requirements presented is related to connection and database operations. Among the highlighted words (in bold), one can find, for example "sqoop". This word is part of to an application that transfers data between relational databases and \textit{Hadoop} \footnote{Hadoop is an open source software platform for the storage and distributed processing of large databases. The services \textit{Hadoop} provides includes storage, processing, access, governance, security and data operations, making use of powerful hardware architectures, which are usually on loan.}. Therefore, the identification of similar contexts is more clearly perceived, even with very different words, as is the case of "sqoop", "sqlserver" and "persistent". These are different words, but part of the same context. Thus, although the groupings did not demonstrate clear groups, either by project or by effort, the groupings demonstrated, at a certain level, a representation of requirements from the same context, even if from different projects and/or efforts.

\subsection{Effort Estimation Model Settings}

\begin{figure}[!ht]
  \centering
  \includegraphics[width=0.48\textwidth]{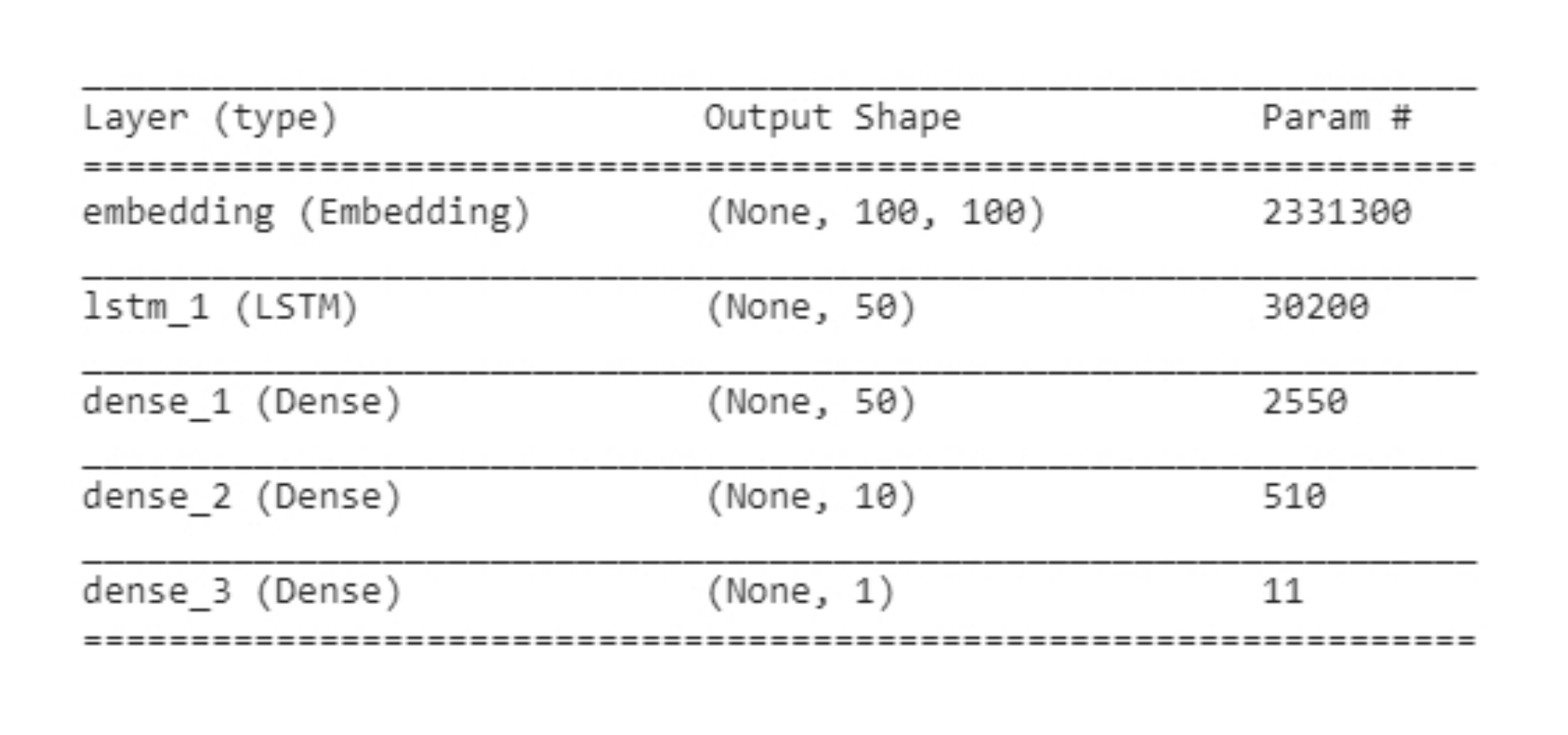}
  \caption{Architecture \textit{deep learning} with the pre-trained embedding layer using Word2Vec.} 
  \label{figure7}
\end{figure}

\begin{figure}[!ht]
  \centering
  \includegraphics[width=0.48\textwidth]{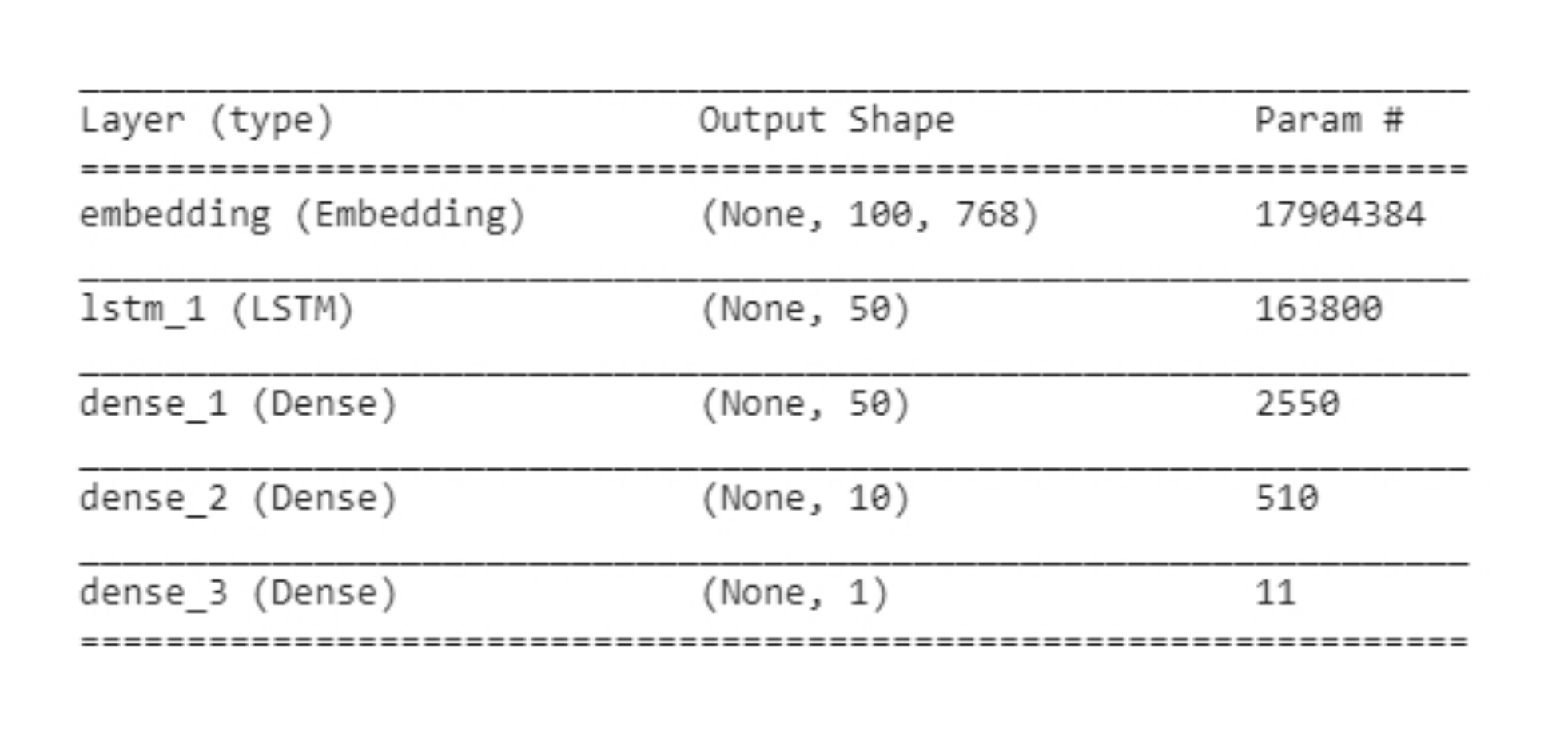}
  \caption{Architecture \textit{deep learning} with the pre-trained embedding layer using BERT.} 
  \label{figure8}
\end{figure}

In this section, the stages of the S$E^3$M model will be presented, in which representation vectors for each textual requirement are obtained according to the procedures presented in section \ref{obtCaract}, are given as a parameters to the layers \textit{dense} of the architecture \textit{deep learning}, as shown in Figures \ref{figure7} and \ref{figure8}.

The \textit{embedding()} layers (Figures \ref{figure7} and \ref{figure8}) are represented by the vector of pre-trained weights for each sentence. These vectors are processed by two dense nonlinear layers, followed by a linear regression layer. The output is the estimate of the predicted effort (e.g. points per story).

Each instance of textual requirement submitted to the \textit{Embedding ()} layer is represented by an average vector representing each sentence. A sentence consists of a maximum of 100 words, each of which is represented by 100 text embeddings for the \textit{Word2Vec} models and 768 dimension embeddings according to BERT models, as justified below.

The maximum number of words per text was defined based on the representation of the histogram shown in Figure \ref{figure2}, which shows that this number would include most of the sentences used in the training and test database, with reduced data loss.

As presented in the architecture specification S$E^3$M (section \ref{const_approach}), it is a very simple architecture, post layer of embedding, consisting only of two dense non-linear layers and a linear regression layer. The fact that the \textit{deep learning} architecture is very simplified was purposeful, considering that the objective of this thesis is to present how to infer effort estimates in software projects by analogy using pre-trained embeddings models, verifying whether these models are promising for text-based software effort estimation. In this way, a more robust architecture could mask the results generated at each of the different network entrances.

A textual requirement, also referred to in this study as a sentence, is represented by an average vector of the word embeddings that compose this sentence. This average vector is generated from each of the generated models (Tables \ref{tab:pret_model} and \ref{tab:model_ajuste}, considering the particularities of each applied approach (without context and contextualized), as presented in section \ref{obtCaract}. Therefore, each generated embedding model is represented as a matrix, where each line represents an average embedding vector for a given textual requirement. Thus, each embedding model has the same number of samples, and what varies is the dimension applied.

Previous tests were performed using the \textit{grid search} method. For the embeddings models without a context, tests were performed with dimensions 50, 100, and 200 with the best results presented by dimensions 50 and 100. As there was no significant variation for the MAE between both dimensions using the same neural network architecture, the number of dimensions of embeddings was fixed at 100. For the \textit{transformers} BERT the standard dimension defined for the \textit{BERT\_base} model was used, that is, equal to 768.

For the training of the learning model, it was necessary to configure some hyperparameters. Therefore, the \textit{Adam} optimizer \cite {kingma2014adam}, learning rate 0.002 was used and the size of the \textit{batch\_size} was 128. Twenty epochs and an early stopping mechanism were defined, in which, if the MAE value remains stable for 5 epochs, the best results are averaged.

\subsection{Experiment Settings}

The results obtained in this study will be presented below in order to make a comparative analysis with the most similar study \cite{choetkiertikul2018deep} we identified, in addition to answering the following research questions defined initially.

\begin{table*}[ht]
    
    \footnotesize
    \begin{tabular}{c|c|c}
      \hline
      \textbf{Experiments} & \textbf{Pre-trained model} & \textbf{Sequential network architecture}   \\
      \hline
      E1 & \textit{word2vec\_base} & Embedding() + LSTM + Dense (not-linear) + Dense (not-linear) + Dense (linear) \\
      \hline
      E2 & \textit{word2vec\_SE} & Embedding() + LSTM + Dense (not-linear) + Dense (not-linear) + Dense (linear) \\
      \hline
      E3 & \textit{BERT\_base} & Embedding() + LSTM + Dense (not-linear) + Dense (not-linear) + Dense (linear) \\
      \hline
      E4 & \textit{BERT\_SE} & Embedding() + LSTM + Dense (not-linear) + Dense (not-linear) + Dense (linear) \\
      \hline
      E5 & \textit{BERT\_SE} & Embedding() + LSTM + Dense (not-linear) + Dense (not-linear) + Dense (softmax) \\
      \hline
    \end{tabular}
    
    \caption{Description of performed experiments.}
    \label{tab:ord_experiments}
\end{table*}

\section{Results and Discussions}

The results will be presented below, to make a comparative analysis with the most similar study \cite{choetkiertikul2018deep}, in addition to answering the following research questions defined initially.

\textbf{RQ1. Does a generically pre-trained word embedding model show similar results to a software engineering pre-trained model?}

To answer this question, experiments (\textbf{E1}, \textbf{E2}, \textbf{E3} and \textbf{E4}) were performed, with pre-trained models with and without fine-tuning, for context-less and contextualized approaches, a task that aimed to determine effort estimation. The approach consists of using a pre-trained embedding model as the only source of input in the deep learning architecture used.

Table \ref{tab:resultgeral} presents the average values for the MAE, MdAE and MSE measurements obtained after the application of cross-validation \textit{10-fold} during the execution of the proposed sequential architecture, in which each model was used as input for pre-trained embedding available for each of the experiments (according to Table \ref{tab:ord_experiments}).

\begin{table*}[ht]

    \footnotesize 
    \begin{tabular}{c|c|c|c|c|c}
        \cline{1-1}\cline{2-2}\cline{3-3}\cline{4-4}\cline{5-5}  
        \multicolumn{1}{c|}{\textbf{Approach} \centering } &
        \multicolumn{1}{c|}{\textbf{Pre-trained model} \centering } &
        \multicolumn{1}{c|}{\textbf{MAE} \centering } &
        \multicolumn{1}{c|}{\textbf{MSE} \centering } &
        \multicolumn{1}{c}{\textbf{MdAE} \centering }\\  
          \cline{1-1}\cline{2-2}\cline{3-3}\cline{4-4}\cline{5-5}  
            \multicolumn{1}{c}{Context-less} &
            \multicolumn{1}{|c|}{word2vec\_base} &
            \multicolumn{1}{c|}{4.66 $\pm$0.14} &
            \multicolumn{1}{c|}{100.26 $\pm$7.04} &
            \multicolumn{1}{c}{2.9}\\  
          
          \cline{2-2}\cline{3-3}\cline{4-4}\cline{5-5}  
            \multicolumn{1}{c|}{ } &
            \multicolumn{1}{c|}{word2vec\_SE} &
            \multicolumn{1}{c|}{4.36 $\pm$0.31} &
            \multicolumn{1}{c|}{89.9 $\pm$14.37} &
            \multicolumn{1}{c}{2.5}\\  
        \cline{1-1}\cline{2-2}\cline{3-3}\cline{4-4}\cline{5-5}  
            \multicolumn{1}{c}{Contextualized} &
            \multicolumn{1}{|c|}{BERT\_base} &
            \multicolumn{1}{c|}{4.52 $\pm$0.094} &
            \multicolumn{1}{c|}{100.95 $\pm$7.3} &
            \multicolumn{1}{c}{2.7}\\  
          
          \cline{2-2}\cline{3-3}\cline{4-4}\cline{5-5}
            \multicolumn{1}{c|}{ } &
            \multicolumn{1}{c|}{BERT\_SE} &
            \multicolumn{1}{c|}{\textbf{4.25 $\pm$0.17}} &
            \multicolumn{1}{c|}{\textbf{86.15 $\pm$1.66}} & 
            \multicolumn{1}{c}{\textbf{2.3}} \\   
        
        \hline
    \end{tabular}
    \caption{Evaluation of the results obtained for experiments E1, E2, E3 and E4. Bold are the best results for each pre-trained embedding model. For all the metrics used, the lower the value, the better the result.}
    \label{tab:resultgeral}
\end{table*}

As can be seen in Table \ref{tab:resultgeral}, the models that underwent fine-tuning (with SE suffix), regardless of the approach used, gave better results for MAE, MSE and MdAE. Thus, it is clear that a pre-trained embedding model, in which fine-tuning is performed with a specific corpus of the task domain, presents a better performance than a pre-trained model with a generic corpus, as is the case with \textit{word2vec\_base} and \textit{BERT\_base}.

This can be proven by comparing the generic context-less embedding model (\textit{word2vec\_base}) to the same fine-tuned model (\textit{word2vec\_SE}), where a 5\% improvement is seen in relation to the MAE value for the second model. Likewise, when applying the contextualized embedding model with fine-tuning (\textit{BERT\_SE}), an improvement of 7.8\% for the MAE was observed in relation to the generic model \textit{BERT\_base}. Thus, it is noted that, in general, the results achieved improved after adjusting the models with a specific corpus of the domain .

If the MAE value obtained for the best model in Table \ref{tab:resultgeral} (\textit{BERT\_SE}), which was 4.03, is applied in practice, this will indicate that, for a given user story, in if the real effort is 5 points per story, the estimated effort value could be between 1 and 9 points per story. This is one of the reasons why human participation in the estimation process is indicated, in order to calibrate these effort values to new requirements.

In terms of MSE, the improvement was of 3.5\% for the \textit{word2vec\_SE} model, when compared to \textit{word2vec\_base}, and 13.5\% for MdAE. When evaluating MSE and MdAE for the contextualized approach, the fine-tuned model outperforms the generic model by 14.6\% and 14.8\%, respectively.

Thus, we conclude that the representation of the training and test model (\textit{corp\_SE}), when generated from pre-trained and adjusted embeddings models, improves the performance of activities such as the ABSEE. To this end, it is estimated that the greater the volume and diversity of samples in the corpus used in fine-tuning (\textit{corpPret\_SE}, the better the results can be. Therefore, this fact must be considered when giving continuity to this task in future studies.

\textbf{RQ2. Would embedding models generated by context-less methods (i.e. Word2Vec) be effective as models generated by contextualized methods (i.e. BERT)?}

As stated by Ruder \cite{howard2018universal}, “it only seems to be a question of time until pre-trained word embeddings (i.e. word2vec and similar) will be dethroned and replaced by pre-trained language models (i.e. BERT) in the toolbox of every NLP practitioner.” Thus, the objective of this study was to analyze if context-less embeddings models are effective in effort estimates, as compared to contextualized embeddings models in a specific corpus.

As can be seen in Table \ref{tab:resultgeral} the results obtained by the contextualized models (\textit{BERT\_base} and \textit{BERT\_ES}), surpass the results of the models without context.

When comparing MAE values, the \textit{BERT\_base} model shows a 3\% improvement over \textit{word2vec\_base}. Likewise, for the values of MdAE and MSE there was an improvement of 0.7\% and 6.9\%, respectively.

When comparing how these metrics between the models with fine tuning (\textit {Word2vec\_ES} and \textit {BERT\_ES}), as improvements of the contextualized model in relation to no context were 2.5\%, 4.2\% and 8\% for the values of MAE, MSE and MdAE, respectively.

It is believed that this result is due to the fact that methods context-less (ex. \textit{Word2Vec}) allow to represent a single context for a given word in a set of texts. This aspect causes a lot of information to be lost. In an effort estimate, based on the requirements texts (e.g. user stories), a contextualized strategy certainly produces better results. Unlike the \textit{Word2vec} approach, BERT methods offer this dynamic context, allowing the actual contexts of each word to be represented in each text.

This aspect is important, as textual software requirements (i.e. user stories, use cases) generally have short texts and little vocabulary, which means that many words are common to the field of software engineering and are repeated in many texts. This is the importance of identifying different contexts of use for each word, in order to differentiate them. Contextualized methods like BERT guarantee this dynamic treatment of each word, addressing problems of polysemy and ambiguity in an intrinsic way.

This is possible due to the fact that the contextualized models use a model of deep representation of each word in the text (that is, 12 or 24 layers), unlike the models of embeddings context-less that present a superficial representation (of a single layer) for a word. In addition, contextualized models use an attention model that allows verifying whether the same word occurred previously in the same context or not (example in Table \ref{tab:exe_poli}), or whether different words can present the same context (e.g. create, implement, generate) - example in Table \ref{tab:exe_ambi}.

\begin{table}[ht]
    
    \footnotesize
    \begin{tabularx}{0.48\textwidth}{l|X}
      \hline
      \textbf{Text 22810} & \textbf{Add} that contact as a favorite notice that the images for contacts (driven by remote url) change unexpectedly. Under the covers all that is happening, is that the data of the list view is refreshed. \\
      \hline
      \textbf{Text 23227} & \textbf{Add} qparam to skip retrieving metadata and graph edges if the qparam is not there, use current default behavior. \\
      \hline
    \end{tabularx}
    
    \caption{Example of requirement texts that have polysemy - same words and different contexts.}
    \label{tab:exe_poli}
\end{table}

\begin{table}[!ht] 
    
    \footnotesize
    \begin{tabularx}{0.48\textwidth}{l|X}
      \hline
      \textbf{Text 18} & {\textbf{create} new titanium studio splash screen there is a placeholder image...} \\
      \hline
      \textbf{Text 22810} & {\textbf{build} the corporate directory app for ios...} \\
      \hline
    \end{tabularx}
    
    \caption{Example of requirement texts that have ambiguity - different words in the same context.}
    \label{tab:exe_ambi}
\end{table}

\textbf{RQ3. Are pre-trained embeddings models useful to a text-based software effort estimation?}

To answer this question regarding the perspectives of contextualized pre-trained models applied to ABSEE, it is necessary to observe whether MAE, MSE and MdAE obtained good results. As shown in Table \ref{tab:resultgeral}, the best MAE value was 4.25, which means that a software effort of 6 will be predicted between 2 and 10.

When questioning whether this result is good or bad, can be observed that, considering the small number of samples in the training set and tests, the high degree of imbalance between classes, and the high variability of the text, this result is quite adequate. It is precisely due to the perception that, at least currently, there is little data to estimate the effort involved in software development that motivated these researchers to investigate pre-trained embedding models, so as to solve the proposed problem.

\begin{table}[htbp]

    \begin{tabularx}{0.48\textwidth} {l|c|c}
      \hline
      \textbf{Method} & \textit{S$E^3$M} (multi-repositories) & \textit{Deep-SE} (cross-repository)  \\
      \hline
      \textbf{MAE} & 4.25 $\pm$ 0.17 & 3.82 $\pm$ 1.56 \\
      \hline
    \end{tabularx}
    
    \caption{Mean Absolute Error (MAE) obtained for \textit{BERT\_SE} compared to the Deep-SE model \cite{choetkiertikul2018deep}}
    \label{tab:comparaMAE}
\end{table}

When comparing the best MAE results, obtained though the \textit{BERT\_SE} model, with the MAE results given by the Deep-SE \cite{choetkiertikul2018deep} model, the latter of which was the study found to be most similar to the present research, some aspects stand out, as shown below.

As can be seen in Table \ref{tab:comparaMAE}, the MAE obtained by \textit{BERT\_SE} was slightly higher than \textit{Deep-SE}. However, one should note that, to obtain this result, \cite{choetkiertikul2018deep} inferred the effort estimates between projects (e.g. Moodle/Titanium, Mesos/Mule). Therefore, there is a large chance that two projects share a similar context, which would make predictability easier for projects in different contexts. This statement is reflected directly in the standard deviation of the MAE values (e.g. 5.37, 6.36, 5.55, 2.67, 4.24) for the Deep-SE between projects, which is 1.56. One can see that some values of MAE are relatively low (e.g. 2.67), while others are higher (ex. 6.36). This means that there may be a higher variation for the estimates inferred by Deep-SE, which, in the worst case, can cause a requirement whose effort is 7 points per story, to return 1 or 13 points per story. Thus, it is suggested that if the Deep-SE model is applied using a single repository approach, as well as the S$E^3$M, the MAE values may be even higher.

S$E^3$M, on the other hand, uses a single repository approach, that is, all requirements are independent of project or repository, aiming to generalize the model. Thus, although the S$E^3$M MAE is close to the Deep-SE value, the standard deviation obtained is smaller (0.17), or almost nonexistent (ex. 3.87, 4.21, 4.15, 4.12, 3.97, 4.25, 4.01), which can be proven by observing the pattern of MAE values obtained by the model. In this sense, it can be said that the proposed method is more generic and applicable to different problems, demonstrating a greater degree of reliability than Deep-SE. This is mainly due to the ability of the BERT \textit{Transformer} mechanism (attention mechanism) to resolve long-term dependency and “vanishing gradient” \cite{hochreiter1991untersuchungen} and \cite{bengio1994learning}, that presents itself as limitations in recurrent network architectures, as is the case with RHN and LSTM, used by \cite{choetkiertikul2018deep}. The “vanishing gradient” is the loss of relevant context information, used to identify the semantics of a given word in a text. Therefore, the attention mechanism allows for one to ignore irrelevant information and focus on what is relevant, making it possible to connect two related words, even if they are not located one after the other.

In order to reinforce the positive trend of applying pre-trained embeddings models in the process of inferring effort estimation, the \textbf{E5} experiment was performed. In this case, the same set of training and testing data was used, applying a classification layer (with softmax activation) instead of linear regression. This required some modifications to the dataset. First, there was a need to make the dataset more homogeneous and bulky as compared to the existing labels. Then the closest estimates were grouped, considering the series of \textit{Fibonacci} proposed for the \textit{Planning Poker} \cite{cohn2005agile} estimation method. As a result of this process, the data set only had 9 possible labels: 1, 2, 3, 5, 8, 13, 20, 40, 100. Therefore, compared to the regression problem, each label had its data volume increased and balanced in relation to the existing classes, mainly for the smaller labels (between 1 and 8), as shown in Table \ref{numreqPP}.

\begin{table*}[!ht]
    \footnotesize
    \begin{tabular}
    {p{0.25\linewidth}|p{0.04\linewidth}|p{0.04\linewidth}|p{0.04\linewidth}|p{0.04\linewidth}|p{0.04\linewidth}|p{0.04\linewidth}|p{0.04\linewidth}|p{0.04\linewidth}|p{0.04\linewidth}|p{0.04\linewidth}}
      \hline
      \textbf{Number of labels} & \textbf{1} & \textbf{2} & \textbf{3} & \textbf{5} & \textbf{8} & \textbf{13} & \textbf{20} & \textbf{40} & \textbf{100} & \textbf{Total}\\
      \hline
      \textbf{Number of textual requirements/labels} &  4225 & 3406 & 4809 & 4725 & 3588 & 1238 & 706 & 451 & 165 & \textbf{23.313}\\
      \hline
    \end{tabular}
    
    \caption{Number of textual requirements allocated to each Planning Poker class.}
    \label{numreqPP}
\end{table*}

\begin{figure}[!htb]
\centering
\includegraphics[width=0.48\textwidth]{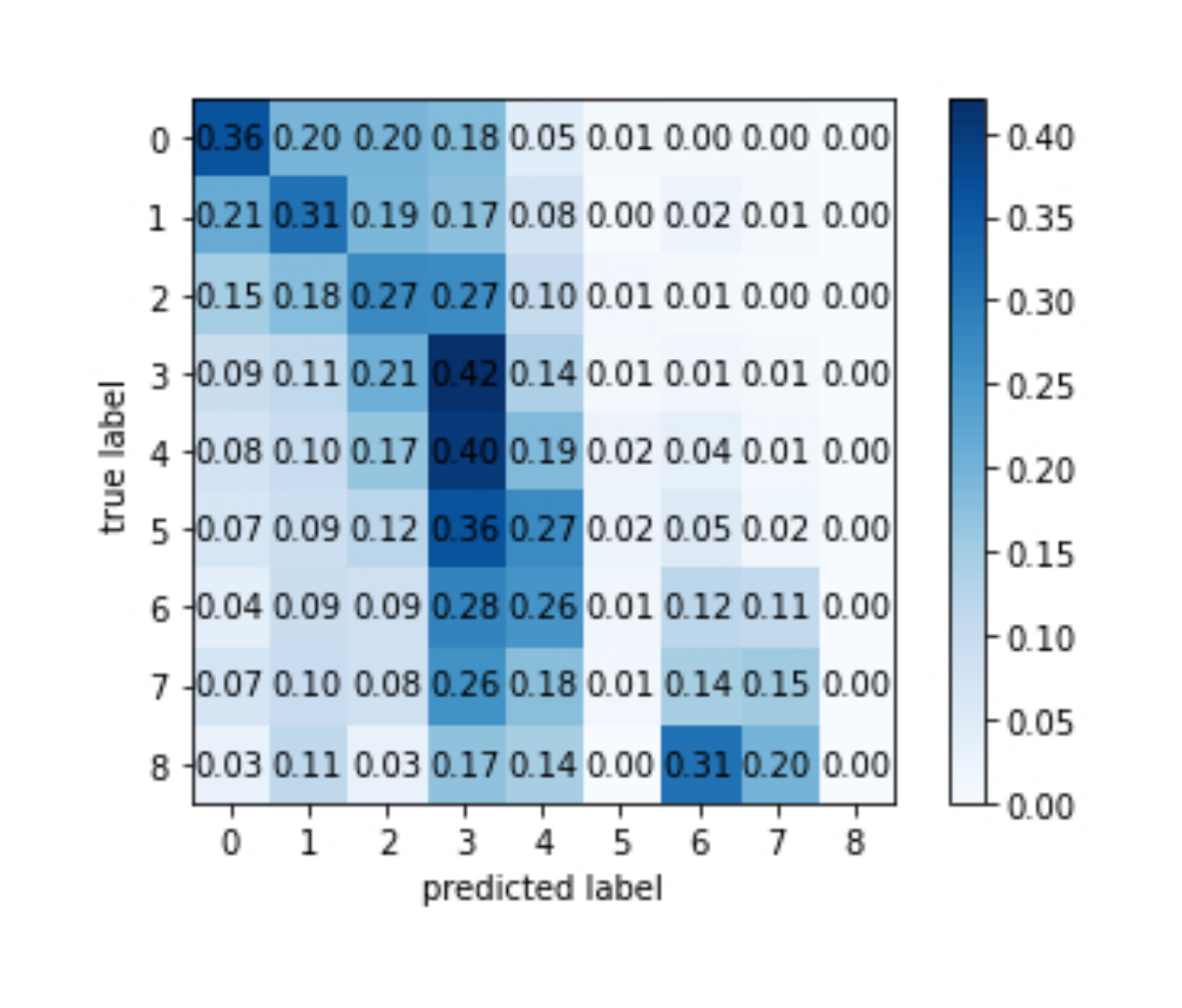}
\caption{Confusion matrix for the E5 experiment using the 9 classes from \textit{Planning Poker}.}
\label{fig_confusion_matrix}
\end{figure}

The Figure \ref{fig_confusion_matrix} shows the confusion matrix generated for the experiment \textbf{E5}. One can see that the greatest confusion occurs between the lowest efforts (1, 2, 3, 5 and 8). Considering the MAE value of the regression experiment (4.25), it is possible to understand this bias in the confusion matrix; after all, as explained previously, an effort of 4 points per story could be 1 or 8.

One can also observe that, similarly, the larger classes (13, 20, 40 and 100) became more confused with each other and, in very low percentages there was confusion among the smaller classes. This aspect leads us to suggest that human intervention at the end of the process is needed, with the aim of approving/modifying the estimate generated, according to the user's working reality. Thus, considering its application for the end user, the proposed method would be classified as semi-automated.

Another aspect to be observed in the results of Figure \ref{fig_confusion_matrix} is an indication that the larger the data set representing each of the labels, the better the results. This study argues that effective techniques for increasing data in texts can improve this result. Another alternative would be to collect a more significant number of real samples, complementing the pre-training and fine-tuning data set \textit{corpPret\_SE}, as well as sets for training and testing (\textit{corp\_SE}). Thus, the generated embeddings will be more representative, that is, they can better represent each situation found in the requirements.

\textbf{RQ4. Are pre-trained embeddings models useful to a text-based software effort estimation, both on new and existing projects?}

Another aspect that can be observed is the indication of the S$E^3$M model to estimates new and non-existing requirements. For this, an additional experiment was carried out containing the same configuration as \textbf{E4}, and only changing the type of data partitioning. In this experiment \textit{cross validation} was applied by projects in order to evaluate the results of inferring the estimates for each project, that is, for completely new projects. Thus, during each of the iterations one of the projects was considered a target for the estimates (test set), while the others were considered a source (training set). Table \ref{tab:crossval_proj} shows the results obtained for the MAE in each project.

\begin{table}[!ht]
    
    \footnotesize
    \begin{tabularx}{0.48\textwidth}{c|c|X|c|c|c}
      \hline
      \textbf{ID project} & \textbf{Description} & \textbf{Num. requirement} & \textbf{Med} & \textbf{Std} & \textbf{MAE}\\
      \hline
      {AS} & {Appcelerator Studio} & {2919} & {5.63} & {3.32} & {2.50}\\
      {AP} & {Aptana Studio} & {829} & {8.01} & {5.95} & {4.18}\\
      {BB} & {Bamboo} & {521} & {2.41} & {2.14} & {2.76}\\
      {CV} & {Clover} & {384} & {4.6} & {6.54} & {3.87}\\
      {DM} & {Data Management} & {4667} & {9.56} & {16.6} & {7.78}\\
      {DC} & {DuraCloud} & {666} & {2.12} & {2.03} & {3.79}\\
      {JI} & {JIRA Software} & {352} & {4.43} & {3.51} & {3.13}\\
      {ME} & {Mesos} & {1680} & {3.08} & {2.42} & {3.39}\\
      {MD} & {Moodle} & {1166} & {15.54} & {21.63} & {11.99}\\
      {MU} & {Mule} & {889} & {5.08} & {3.49} & {3.51}\\
      {MS} & {Mule Studio} & {732} & {6.39} & {5.38} & {3.51}\\
      {XD} & {Spring XD} & {3526} & {3.69} & {3.22} & {3.16}\\
      {TD} & {Talend Data Quality} & {1381} & {5.92} & {5.19} & {4.04}\\
      {TE} & {Talend ESB} & {868} & {2.16} & {1.49} & {3.42}\\
      {TI} & {Titanium SDK/CLI} & {2251} & {6.31} & {5.09} & {3.49}\\
      {UG} & {Usergrid} & {482} & {2.85} & {1.40} & {3.24}\\
      \hline
    \end{tabularx}
    
    \caption{MAE values when estimating the effort for each project in relation to the others. The number of requirements by project (Num. requirement), the average (med) and standard deviation (std) of the effort by requirement in each project are presented.}
    \label{tab:crossval_proj}
\end{table}

When considering the cross-project estimation approach, if target projects are considered (UG, ME, AP, TI, AS, TI, MS, MU), as listed in Table 9 of the article by \cite{choetkiertikul2018deep}, the authors were able to an average MAE value of 3.82 for the Deep-SE model, while the S$E^3$M model presented a MAE of 3.4 (Table \ref{tab:MAE_reqNovos}). It is observed that this comparison was made only with the results obtained in the mode between repositories of \cite{choetkiertikul2018deep}, and for the results obtained with S$E^3$M, all 15 projects were considered as source projects remaining, while the target projects are the same chosen by the authors.

It is still observed that the authors of Deep-SE used a source project for training and another target project for tests, that is, the diversity of characteristics that the learning method needs to deal with, is limited only by the domain of a single project. In other words, the model needs to deal with less variability of data, which supposedly can facilitate learning, since some important relationships between these data can be discovered more easily by the learning method. However, the proposal presented here uses all projects (except one) for training, that is, the variety of relationships that the model must deal with is much greater, when compared to the Deep-SE approach. From a certain point of view this is good, as the learning method should learn a better generalization, for any type of problem presented. On the other hand, however, learning is hampered due to curse of dimensionality, in which many important relationships can be more difficult to be inferred automatically, mainly due to the very small corpus. The explanation for this is that in a smaller data set, basic relationships are easily learned, whereas more specific relationships are often not sufficiently representative.

In the case of software projects, each project used as training and testing data is in fact an another domain (when compared to another project), which can be composed of several sub-domains, such as: application areas of the project, programming languages (e.g. Java, Python, C), databases (e.g. SQL Server, Oracle, MySQL, MongoDB), application modalities (e.g. web, mobile, desktop), development teams with different characteristics (e.g. beginner, full), among others. Therefore, each sub-domain can be composed of different relationships, since different aspects can change the context of one project in relation to the other, which leads, for example, to the use of different terms for the same purpose. Thus, there is a need for a balanced volume of representative samples from each sub-domain, so that the model can learn properly.

\begin{table}[!ht] 

    \footnotesize
    \begin{tabularx}{0.48\textwidth}{c|c|c|c}
      \hline
      \textbf{Source projects} & \textbf{Target projects} & \textbf{MAE (Deep-SE)} &  \textbf{MAE (S$E^3$M)}\\
      \hline
      {} & {UG} & {1.57} & {3.24} \\
      {} & {ME} & {2.08} & {3.39} \\
      {} & {AP} & {5.37} & {4.18} \\
      {15 remaining projects} & {TI} & {6.36} & {3.49} \\
      {} & {AS} & {5.55} & {2.50} \\
      {} & {TI} & {2.67} & {3.49} \\
      {} & {MS} & {4.24} & {3.51} \\  
      {} & {MU} & {2.70} & {3.51} \\
      \hline
      \textbf{Avg} & {} & {3.82} & \textbf{3.40} \\
      \hline
    \end{tabularx}
    
    \caption{Mean Absolute Error (MAE) in estimating effort for new projects.}
    \label{tab:MAE_reqNovos}
\end{table}

Although the results of this study are not very different from that obtained in the article by \cite{choetkiertikul2018deep}, the proposed approach presents a much simpler and potentially more robust network architecture. This is possible because part of the textit{feature learning} process previously performed, which extracts the contextualized textual representation for new project requirements, is performed only once, during the process of generating the pre-trained model (\textit{BERT\_SE}). This model is then employed as a parameter in the embedding layer of the sequential architecture used. This aspect is important, considering that there is the need to feed the training database with new cases of requirements, as well as with the cases’ respective efforts, which makes it possible to increase precision in effort estimates for new projects.

\textbf{RQ5. Are the results found generalizable, aiming to generate estimates of effort between companies?}

In our approach, the results show that even a new project can have its effort predicted without any pre-existing data (as shown in RQ4). Although the MAE still does not deliver a perfect result, a good estimate of the effort can be achieved and used, in a semi-automated way, by companies, as explained in the RQ3 response.

Additionally, when considering the application of the model for multiple companies, it is necessary to consider the possibility that existing requirements contain different metrics (e.g. function points, story points), since the data-set would be fed by requirements from different companies. To address this question, the means of converting these metrics into a standard form, which, if obtained from the estimator, could be converted into the format to be used by the user, is suggested. This conversion of software effort metrics is proposed by \cite{pressman2016engenharia}. 

Thus, generalizing the proposed method so that is can be used by several companies in a web application, for example, would be perfectly possible. This claim is supported by the fact that the method is based on a single repository (i. e. grouping multiple repositories), in which different projects’ requirements are met, regardless of the format for registering the requirements texts.

Furthermore, considering its practical application, the proposed model can be adjusted by the user. Therefore, real estimates generated and approved by specialists can be fed back, making the system more and more adjusted to a specific company, for example.

\section{Threats to validity}

Thus as several research in the area of software engineering that apply machine learning techniques for the use with the texts, this paper propose a kind of software estimate or effort by analogy from requirements texts, difficult approaches regarding the availability of real data. These data should reflect the reality of software projects in different areas, levels of complexity, uses of technologies, among other attributes. Typically, these difficulties are related to the volume of data, the language in which the data is given, the quality of the data (for example, text formats, completed, nonexistent labels, among others). This way, after accomplished a search for textual requirements databases, we decided to use database used by \cite{choetkiertikul2018deep}, which provided the first database at the level of requirements for the realization of area research. Since the model proposed in this article (S$E^3$M) aims to compare the results obtained for estimating software effort with the Deep-SE model, proposed by \cite{choetkiertikul2018deep}, it was defined by using same database.

Therefore, actions to containing or reducing threats to validity carried out by \cite{choetkiertikul2018deep}, were adopted and maintained for this work. Are they:

\begin{itemize}

\item It is used the Actual project requirements data, which were obtained from large and different open source projects.

\item  The story of points per user, that accompany each textual requirement, were first estimated by human teams, and therefore, may not be accurate in some situations. \cite{choetkiertikul2018deep} performed two tests to mitigate this threat: one with the original story points, the other with normalized and adjusted story points. With that it was verified that the proportions of points for history attributed to each requirement were adequate.

\item It was observed that project managers and analysts determine the estimate for a new requirement, based on their comparison with requirements already implemented in the past, and thus carry out the estimate consistently. In this way, the problem is indicated for a machine apprentice, since the training and testing database presents the description of the requirements, accompanied by their respective efforts in points by history. Thus, new requirements can be estimated, as long as they have these two attributes. 

\end{itemize}

In order to perform the experiments, appropriate metrics were applied to evaluate regression models, which are commonly used to evaluate models of software effort estimates \cite{choetkiertikul2018deep} which attest to the validity of the model. In addition, different forms of data partitioning (\textit{cross-validation}) were used, in order to validate the results obtained.

In order to compare the results obtained in this work with those obtained by \cite{choetkiertikul2018deep}, considering that our implementation may not present all the details that Deep-SE presents, our model was tested using the same data set as the authors. Thus, it was possible to state that our results are consistent.

In order to validate the possibility of generalization, it should be noted that the training and testing data set is composed of 23.313 requirements from sixteen open source projects, which differ significantly in size, complexity, developer team and community \cite{choetkiertikul2018deep}. It is observed that open source projects do not present the same aspects as commercial projects in general, especially in relation to the human resources involved, which requires more research. It would be prudent to test the S$E^3$ M model with a database of commercial projects only (with data available containing the same attributes used in this research) and then with all types of projects (commercial and open source) in order to check for significant differences.

\section{Conclusions and Future Work}

The main objective of this research is to evaluate if pre-trained embeddings models are promising for the inference of text-based software effort estimation, evaluating two approaches to embeddings: context-less and contextualized. The study obtained positive results for the pre-trained models for the ABSEE task, particularly the contextualized models, such as BERT. As predicted in the literature, the contextualized methods demonstrate the best performance numbers. In addition, we show that fine-tuning the embedding layers can help improve the results. All of these results can be improved, especially if trained with more data and/or using some effective data augmentation.

The researchers observed that the database was a limitation of this research, particularly because it is not very bulky, which prevents the results from being even better, especially when using cutting-edge PLN techniques (e.g. pre-trained models, fine-tuning and deep learning). When observing the volume of data used in fine-tuning tasks for specific domains, such as \cite{beltagy2019scibert} and \cite{lee2020biobert}, it is clear that even though the domain’s base is considered to be light, they contain billions of words. On the other hand, the domain-specific database applied in the fine-tuning of the \textit{BERT\_SE} model, has around 800 thousand words. Thus, it was observed that the results obtained with the use of BERT could be improved if there was a more voluminous and diverse set of data, in which it was possible to better adjust the model for different problems, or even train a model of its own (from scratch), which would present an even greater level of adjustment according to the domain area.

Thus, this study argues that the S$E^3E$M model analyzed in this article can best adapt to different project contexts (for example, agile development). Estimation of story points or another similar estimation metric (e.g. use case points or function points) have a fine granularity (i.e. they are assigned to each user requirement). But this same inference method can be applied to estimate a coarser granularity element. An example would be a sprint in agile models \cite{paasivaara2009using}, which is estimated by the sum of the smaller tasks that compose it.

Compared with the results obtained by Choetkiertikul et al. \cite{choetkiertikul2018deep}, the most similar work considering the objectives, note that S$E^3$M used a pre-trained contextualized incorporation layer, which went through a fine tuning process, without the need to add any noise to the texts. In addition, the proposed architecture is more simplified, with just one recurring layer. As a result of the feature learning process, which applies a contextualized incorporation approach, we have a more generalized and multi-project method, which can be applied even in new projects. Thus, the method has more flexibility regarding the format of the input and multi-project texts, allowing for interference to be used in any new requirement, even during the initial stage of development.

Using the pre-trained BERT model (even the generic one), there is no need for prior training of a specific corpus, or one that requires a large volume of data. This has the advantage of involving no pre-training cost (which takes days \cite{howard2018universal}). Rather, there is only the need for fine-tuning, which takes a few hours (\cite{devlin2019bert}). That is, there is no need for training from scratch, as performed by \cite{choetkiertikul2018deep}.

Thus, we provide the pre-trained BERT\_SE model that can be used in various software engineering tasks, in addition to allowing for further adjustments, if necessary. In addition, the S$E^3$M model can be applied in a generic way and, in addition to being reliable, it is a cheap and provides for a computationally fast solution, due to the fine tuning process.

The results demonstrated that this is a promising research field with many available resources and room for innovation. Therefore, several research possibilities are presented as future work:

\begin{itemize}
\item Collect more textual requirement data to balance the dataset against existing labels, and thereby increase the number of contexts; or apply data augmentation techniques to improve results.

\item update BERT\_base vocabulary, including specific vocabulary, and then fine-tune.

\item perform fine tuning, such as with model BERT\_large, and compare the results.

\item study and apply effective data augmentation techniques in order to balance the number of samples in each existing effort class, and thus obtain possible improvements in the results.

\item study and apply different combinations of the layers in the BERT model, in order to evaluate the performance of the model regarding fine-tuning and its effects on EESA.

\item study and apply pre-trained "light" (e.g. ALBERT \cite{lan2019albert}) to the S$E^3$M model, and evaluate its performance aiming to achieve EESA.

\end{itemize}

\bibliographystyle{acm}

\bibliography{output}

\end{document}